\documentclass[preprint]{aastex63}

\usepackage{caption}
\usepackage{subcaption}
\newcommand{\ARIEL}{\textit{Ariel}}

\received{ }
\revised{ }
\accepted{ }
\submitjournal{ApJ}

\shorttitle{Alfnoor Tier 1}
\shortauthors{Mugnai et al.}

\begin{document}

\title{Alfnoor: assessing the information content of \ARIEL's low resolution spectra with planetary population studies.}

\correspondingauthor{Lorenzo V. Mugnai}
\email{lorenzo.mugnai@uniroma1.it}

\author[0000-0002-9007-9802]{Lorenzo V. Mugnai}
\affiliation{Dipartimento di Fisica, La Sapienza Universit\`a di Roma,\\
	 Piazzale Aldo Moro 2, 00185 Roma, Italy}

\author[0000-0003-2241-5330]{Ahmed Al-Refaie}
\affiliation{Department of Physics and Astronomy, University College London,\\
	Gower Street, London, WC1E 6BT, UK}

\author[0000-0002-8846-7961]{Andrea Bocchieri}
\affiliation{Dipartimento di Fisica, La Sapienza Universit\`a di Roma,\\
	 Piazzale Aldo Moro 2, 00185 Roma, Italy}

\author[0000-0001-6516-4493]{Quentin Changeat}
\affiliation{Department of Physics and Astronomy, University College London,\\
	Gower Street, London, WC1E 6BT, UK}

\author[0000-0002-3242-8154]{Enzo Pascale}
\affiliation{Dipartimento di Fisica, La Sapienza Universit\`a di Roma,\\
	Piazzale Aldo Moro 2, 00185 Roma, Italy}

\author[0000-0001-6058-6654]{Giovanna Tinetti}
\affiliation{Department of Physics and Astronomy, University College London,\\
	Gower Street, London, WC1E 6BT, UK}



\begin{abstract}

The \ARIEL\ Space Telescope will provide a large and diverse sample of exoplanet spectra, performing spectroscopic observations of about 1000 exoplanets in the wavelength range $0.5 \to 7.8 \; \mu m$. 
In this paper, we investigate the information content of \ARIEL's Reconnaissance Survey low resolution transmission spectra. Among the goals of the \ARIEL\ Reconnaissance Survey is also to identify planets without molecular features in their atmosphere. In this work, (1) we present a strategy that will allow to select candidate planets to be reobserved in a \ARIEL's higher resolution Tier; (2) we propose a metric to preliminary classify exoplanets by their atmospheric composition without performing an atmospheric retrieval; (3) we introduce the possibility to find other methods to better exploit the data scientific content.

\end{abstract}


\section{Introduction}
\label{sec:intro} 

In the last decade the number of known exoplanets has increased tenfold: at the end of 2009 around 400 exoplanets were known, while at the end of 2019 the confirmed discoveries reached more than 4000. This rapid increase in the exoplanetary science yield is expected to continue and it will affect not only the number of discovered planets, but also our knowledge of planetary formation and evolution. While the discoveries will increase thanks to space missions such as \textit{TESS} \citep{Ricker2016}, \textit{CHEOPS} \citep{Cessa2017}, \textit{PLATO} \citep{Rauer2014} and \textit{GAIA} \citep{GAIA2016}, and to ground instrumentation such as \textit{HARPS} \citep{Mayor2003}, \textit{HATnet} \citep{Bakos2018},  \textit{WASP} \citep{Pollacco2006},  \textit{KELT} \citep{Pepper2018}, \textit{OGLE} \citep{Udalski2015}, \textit{NGTS} \citep{Wheatley2013} and many others, our understanding of planets' histories can only grow through planetary composition analysis.

The most effective strategy used today to reveal the atmospheric chemistry and thermodynamics of transiting exoplanets is to use multi-band photometry and spectroscopy \citep[e.g.][]{Seager2000, Charbonneau2005, Tinetti2007, Sing2016, Madhusudhan2012, Huitson2012,Kreidberg201,Edwards2020,Pluriel2020,Guilluy2021,Mugnai2021A}. 
Current instrumentation has enabled this kind of atmospheric characterisation for a few tens of exoplanets over a limited wavelength range \citep[e.g.][]{Sing2016, Tsiaras2018}.
To interpret the observed spectra, spectral retrieval techniques, often developed for the study of the Earth and Solar System planets, have flourished and were adapted to the new field of investigation \citep[e.g.][]{Irwin2008, Line2013, Waldmann2015, Gandhi2017, Al-Refaie2020}. Most recently, an intense effort has been performed to compare and validate different models developed by different teams to assess potential discrepancies among them \citep{Barstow2020}, demonstrating the robustness and consistency of those models.

The Atmospheric Remote-Sensing Infrared Exoplanet Large-survey, \ARIEL, will enable the spectroscopic observation of a diverse sample of about 1000 exoplanets (\citet{Tinetti2018}; Ariel Definition Study Report\footnote{\url{https://sci.esa.int/web/ariel/-/ariel-definition-study-report-red-book}}) in the $0.5 \to 7.8 \; \mu m$ wavelength range. The \ARIEL\ payload has three photometers (VISPhot, $0.5 \; \mu m - 0.6 \; \mu m$; FGS-1, $0.6 \; \mu m - 0.80 \; \mu m$; FGS-2, $0.80 \; \mu m - 1.1 \; \mu m$) and three spectrometers (NIRSpec, $1.1 \; \mu m - 1.95 \; \mu m$ and $R \geq 15$; AIRS CH0, $1.95 \; \mu m-3.9 \; \mu m$ and $R \geq 100$;  AIRS CH1, $3.9 \; \mu m-7.8 \; \mu m$ and $R \geq 30$).
After each observation, the resulting spectrum from each spectrometer is binned during data analysis to optimise the planetary spectrum Signal-to-Noise-Ratio (SNR). Therefore, implementing different binning options, the mission will adopt a four Tiers strategy, expected to deliver spectra with different SNR to optimise the science return \citep{Tinetti2018}. 

Tier 1 was created to deliver a reconnaissance survey where all planets are first observed at low spectral resolution, and only a subset of Tier 1 planets will be further observed to reach SNR $\geq 7$ at a higher spectral resolution (Tier 2, Tier 3). Tier 1 observations have a SNR $\geq  7$ when raw spectra are binned into a single spectral point in NIRSpec, two in AIRS-CH0 and one in AIRS-CH1, for a total of 4 spectral and 3 photometric data points. For $ \sim 50 \%$ of total observed planets, \ARIEL\ will provide spectra at Tier 2 resolution. In this Tier, raw spectra are binned at, respectively,  $R=10$,  $50$, and $15$ in NIRSpec, AIRS-CH0 and AIRS-CH1, with a SNR of 7 or larger. Tier 3 is meant to provide spectra with SNR $\sim 7$ for $5$ to $10\%$ of the total observed targets. In this Tier the raw spectral data are binned at $R=20, \, 100, \, 30$ in NIRSpec, AIRS-CH0 and AIRS-CH1, respectively. Finally, Tier 4 is conceived for bespoke or phase curves observations. Among the main goals  of Tier 1 observations is that to identify planetary spectra that show no molecular absorption features, and to select those to be re-observed in the successive Tiers.

The aim of this paper is threefold:
\begin{enumerate}
    \item to show the capability of selecting the planets with featureless spectra, that may not be observed again in successive Tiers, without involving retrieval techniques;
    \item to introduce a metric and show its principal applications as a tool to classify Tier 1 observed planets on their molecular content, to aid in the selection of targets to be re-observed in successive Tiers;
    \item to show other strategies to exploit \ARIEL\ Tier 1 data are possible such as those based on Machine learning. 
\end{enumerate}

In Sec. \ref{sec:method} we present our strategy to address these three goals. Our new software, Alfnoor, able to build entire planetary populations is presented in Sec. \ref{sec:alfnoor}. Then we discuss the targets chosen to build the populations and the atmospheric properties used in Sec. \ref{sec:planetary population}. In the same section, we also describe a method to identify the flat spectra in the sample (Sec. \ref{sec:flat_spectra}), which is the first paper goal. This method results are then described in Sec. \ref{sec:flat_results}. Then we describe the metric developed as mentioned in the second goal of this paper (Sec. \ref{sec:metric}) and we introduce a classification algorithm to compare the metric with (Sec. \ref{sec:KNN}). We present in detail the results obtained by our algorithm (Sec. \ref{sec:first_results}), we show the relation between the metric and the input molecular abundances in the planets, and we discuss biases and limitations. 
Finally, we provide a preliminary assessment of the application of Machine and Deep Learning techniques to the problem of  spectra classification in Sec. \ref{sec:deep_learning}, discussing their performance in Sec. \ref{sec:deep_results}, but leaving a more thorough investigation to future work. 
In Sec. \ref{sec:discuss} we discuss and compare the results in details.

\section{Methodology}
\label{sec:method}

\subsection{The Alfnoor software}
\label{sec:alfnoor}
\ARIEL\ will provide a sample of hundreds of planetary spectra. To simulate this data set we develop a new algorithm: Alfnoor, the thousand lights simulator, which was also used for Tier 2 data in \citet{Changeat2020}. Alfnoor is a wrapper of TauREx 3 \citep{Al-Refaie2020} and ArielRad \citep{Mugnai2020}. TauREx 3 is a complete rewrite of the atmospheric retrieval code TauREx \citep{Waldmann2015b, Waldmann2015}.
ArielRad is the \ARIEL\ radiometric model: a software that, given the \ARIEL\ payload and mission strategy descriptions, can simulate the signal propagating from a candidate target through the instruments, and return the expected instrument noise. ArielRad, therefore, can compute the number of observations needed to match each of the \ARIEL\ Tier requirements (to reach a minimum SNR=7 at the Tier spectral resolution). 

By combining the two software, Alfnoor produces the atmospheric high resolution forward model of a planet with TauREx 3, it bins down the spectrum to the \ARIEL\ Tier wavelength grid and adds the expected noise estimated by ArielRad. Consequently, Alfnoor returns a simulation of the planet spectrum as observed in each of the \ARIEL\ mission Tiers. Iterating this procedure for different planets or compositions, Alfnoor automates the process of building entire planetary populations and therefore a data set that is representative of the one \ARIEL\ will provide. 

The Alfnoor and the ArielRad tools are not publicly available, currently. However, both TauREx 3\footnote{\url{https://github.com/ucl-exoplanets/TauREx3_public}}\footnote{\url{https://pypi.org/project/taurex/}} and a generic radiometric simulator called ExoRad 2.0\footnote{\url{https://github.com/ExObsSim/ExoRad2-public}}\footnote{\url{https://pypi.org/project/exorad/}}, are publicly available on GitHub and PyPI. 
ArielRad is ExoRad 2.0 configured for the \ARIEL\ payload. 

\subsection{Planetary populations}
\label{sec:planetary population}

To build a diverse sample of planets in terms of masses, radii and temperatures, we use the \ARIEL\ candidates list of \citet{Edwards2019}. This list contains 1000 planets, selected from both NASA's Exoplanet Archive and TESS predicted discoveries, and covers a wide range of planetary radii (from $\sim 0.4$ to $\sim 27 \, R_{\oplus}$), masses (from $\sim 0.01$ to $\sim 3000\, M_{\oplus}$) and equilibrium temperatures (from $\sim 200\, K$ to $\sim 3900 \, K$). From that list, we extract the parameters listed in Tab. \ref{tab:star_planet_info}. Our goal is not to reproduce accurately the composition of the planets in that list, but to test a diverse sample, and therefore we randomly build an atmosphere for each of the listed targets. We produce three planetary populations that will be of use for this work. We call them POP-I, POP-II and POP-III.

\begin{table}[]
	\centering
	\caption{List of host star and planet information obtained from the \ARIEL\ planets candidate list and used to build the planetary populations used in this work.}
	\label{tab:star_planet_info}
	\begin{tabular}{ll}
		\hline\noalign{\smallskip}
		{\bf \centering Star} & {\bf \centering Planet}\\ 
		\noalign{\smallskip}\hline\noalign{\smallskip}
		mass & mass \\
		radius & radius \\
		effective temperature & equilibrium temperature \\
		distance & distance from the star \\
		& orbital period \\
		& transit duration \\
		\noalign{\smallskip}\hline
	\end{tabular}
	
\end{table}

\paragraph{POP-I}

For each planet we randomise the equilibrium temperature, choosing a value between $0.7 \times T_{p}$ and $1.05 \times T_{p}$, where $T_{p}$ is the planet equilibrium temperature in \citet{Edwards2019}. This randomisation is biased toward lower temperature values as we probe the terminator region, where the spectral features are affected both by the day side and the night side temperatures \citep{Caldas2019,Pluriel2020A,Skaf2020}. The temperature randomisation range is consistent with the work presented in \citet{Changeat2020}.


Then, for each planet we consider an isothermal temperature-pressure profile; we add a constant vertical chemical profile \citep{Moses2011} for every molecule from a list of selected molecules (the abundances are randomised according to defined boundaries). Finally, we add randomly generated grey opaque clouds. We use the plane-parallel approximation, building 100 plane-parallel layers to uniformly sample in log-space the pressure range $10^{-4} \to 10^6 \; \rm{Pa}$. Every atmosphere is built with randomised relative abundances of CH$_4$, H$_2$O, CO$_2$ and  NH$_3$ on a uniform logarithmic scale between $10^{-7}$ and $10^{-2}$. Such a large range allows us to explore the sensitivity of our developed method to very different abundances. We also randomised the cloud surface pressure varying between $5 \times 10^2$ and $ 10^6 \; \rm{Pa}$, similarly to what presented in \citet{Changeat2020}, to explore the whole range from overcast to cloud-free atmospheres respectively. Using these boundaries, we obtain that $\sim 40 \%$ of the atmospheres in the populations contains clouds to at least $10^{4} \; \rm{Pa}$ (surface pressure), as expected from \citet{Tsiaras2018} and \citet{Iyer2016}. Every planet is considered filled with a H$_2$ and He atmosphere with mixed ratio He/H$_2 = 0.17$. A list of the opacities used in this work is reported in Tab. \ref{tab:opacities}. 

As already mentioned, following the aims of this paper, we don't focus on the consistency of the atmospheric models used to build the population. The spectra generated will only be used as ``transmission spectral shapes'' to test our methods against. No information other than the planet transmission spectrum is used in this work.

\begin{table}[]
	\centering
	\caption{List of opacities used in this work and their references.}
	\label{tab:opacities}
	
	\begin{tabular}{ll}
		\hline\noalign{\smallskip}
		{\bf \centering Opacity} & {\bf \centering Reference}\\ 
		\noalign{\smallskip}\hline\noalign{\smallskip}
		H$_2$-H$_2$ & \citet{Abel2011,Fletcher2018} \\
		H$_2$-He & \citet{Abel2012} \\	
		H$_2$O & \citet{Barton2017,Polyansky2018} \\
		CH$_4$ & \citet{Hill2013,Yurchenko2014} \\
		CO$_2$  & \citet{Rothman2010} \\
		NH$_3$  & \citet{Yurchenko2011, Tennyson2012} \\
		\noalign{\smallskip}\hline
	\end{tabular}
\end{table}

\begin{figure}
	\centering
	\begin{subfigure}[b]{0.32\textwidth}
		\centering
		\includegraphics[width=\textwidth]{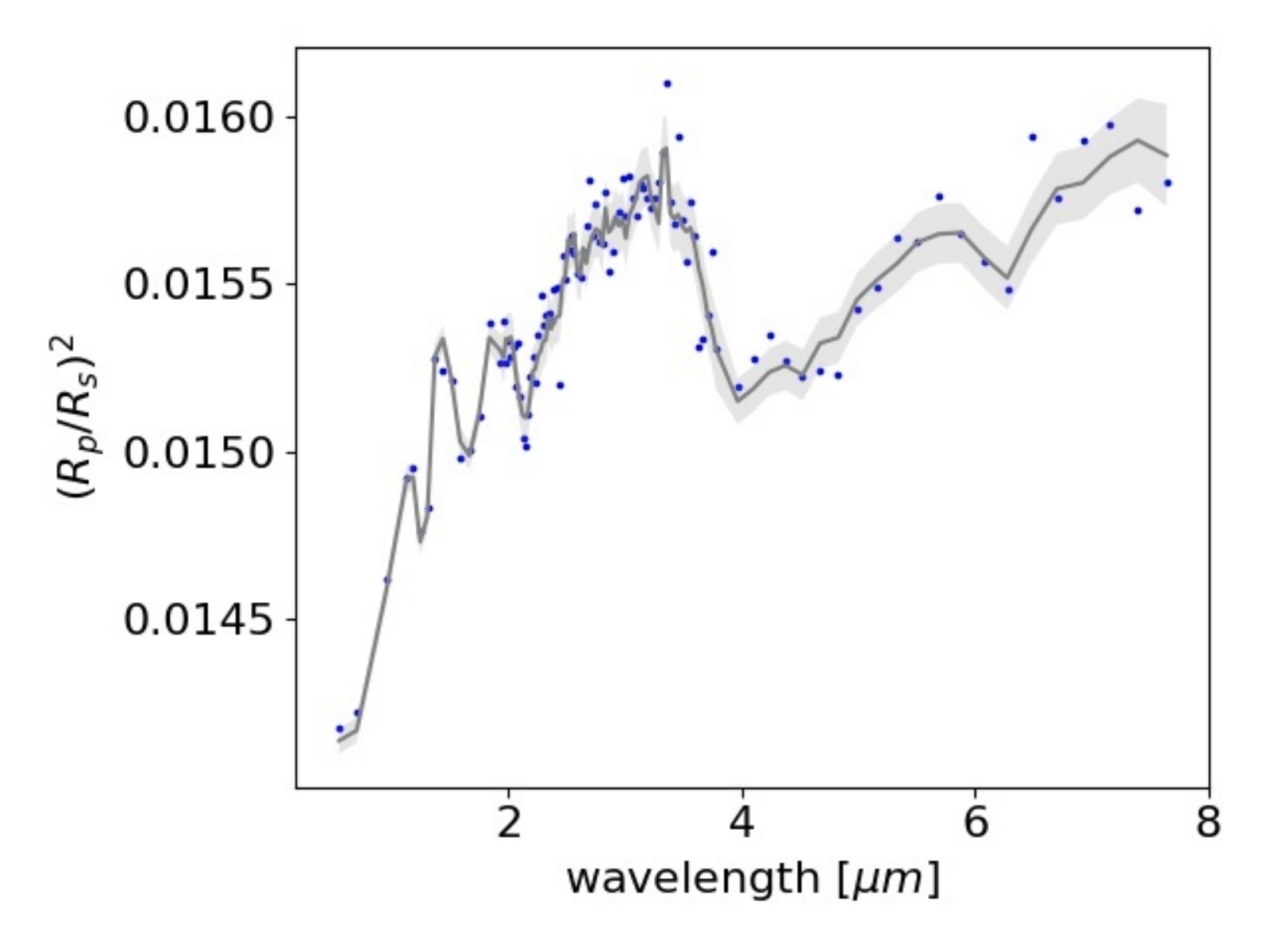}
		\caption{HD 209458b -like planet.}
	\end{subfigure}
	\hfill
	\begin{subfigure}[b]{0.32\textwidth}
		\centering
		\includegraphics[width=\textwidth]{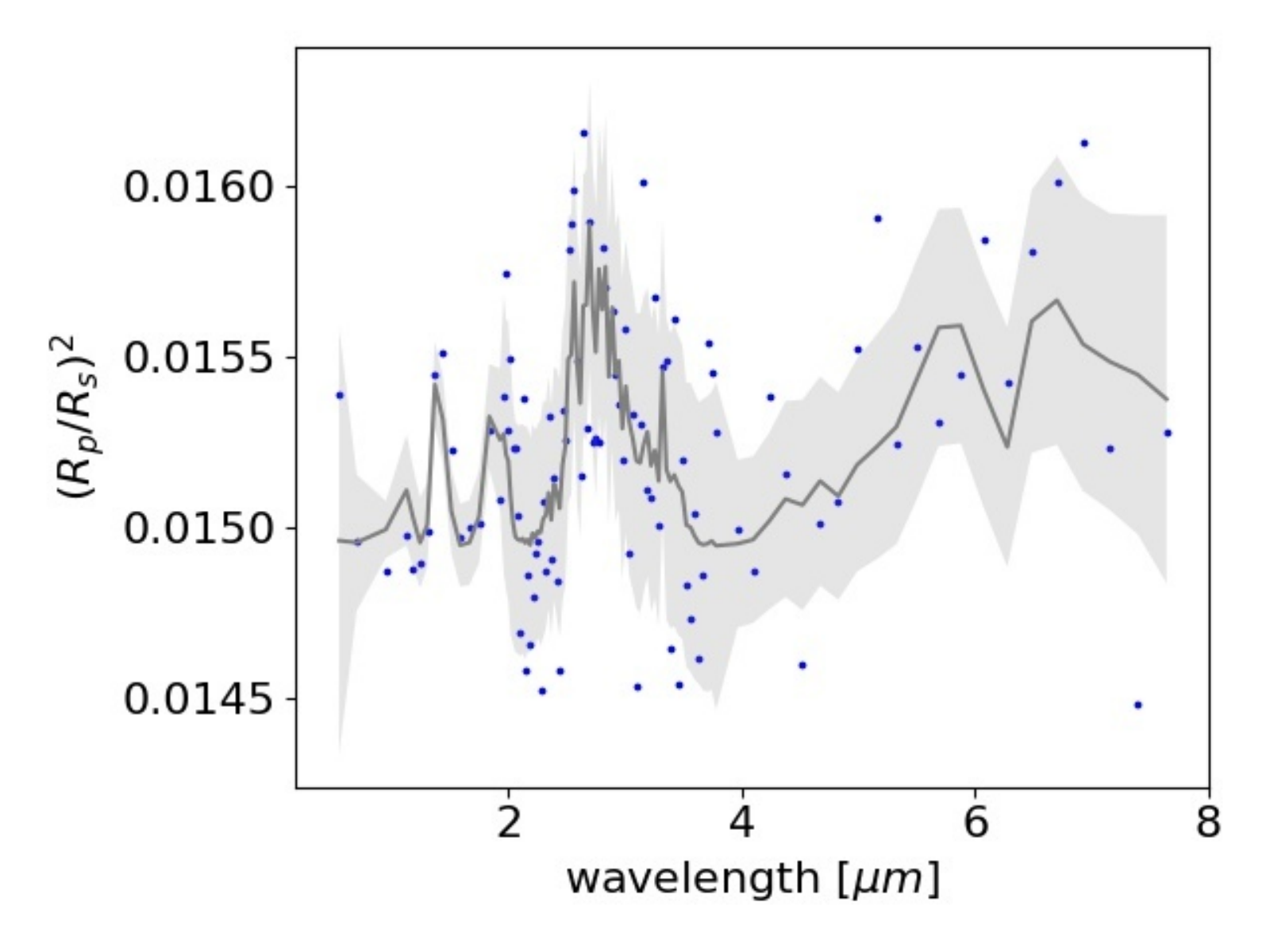}
		\caption{GJ 1214b -like planet.}
	\end{subfigure}
	\hfill
	\begin{subfigure}[b]{0.32\textwidth}
		\centering
		\includegraphics[width=\textwidth]{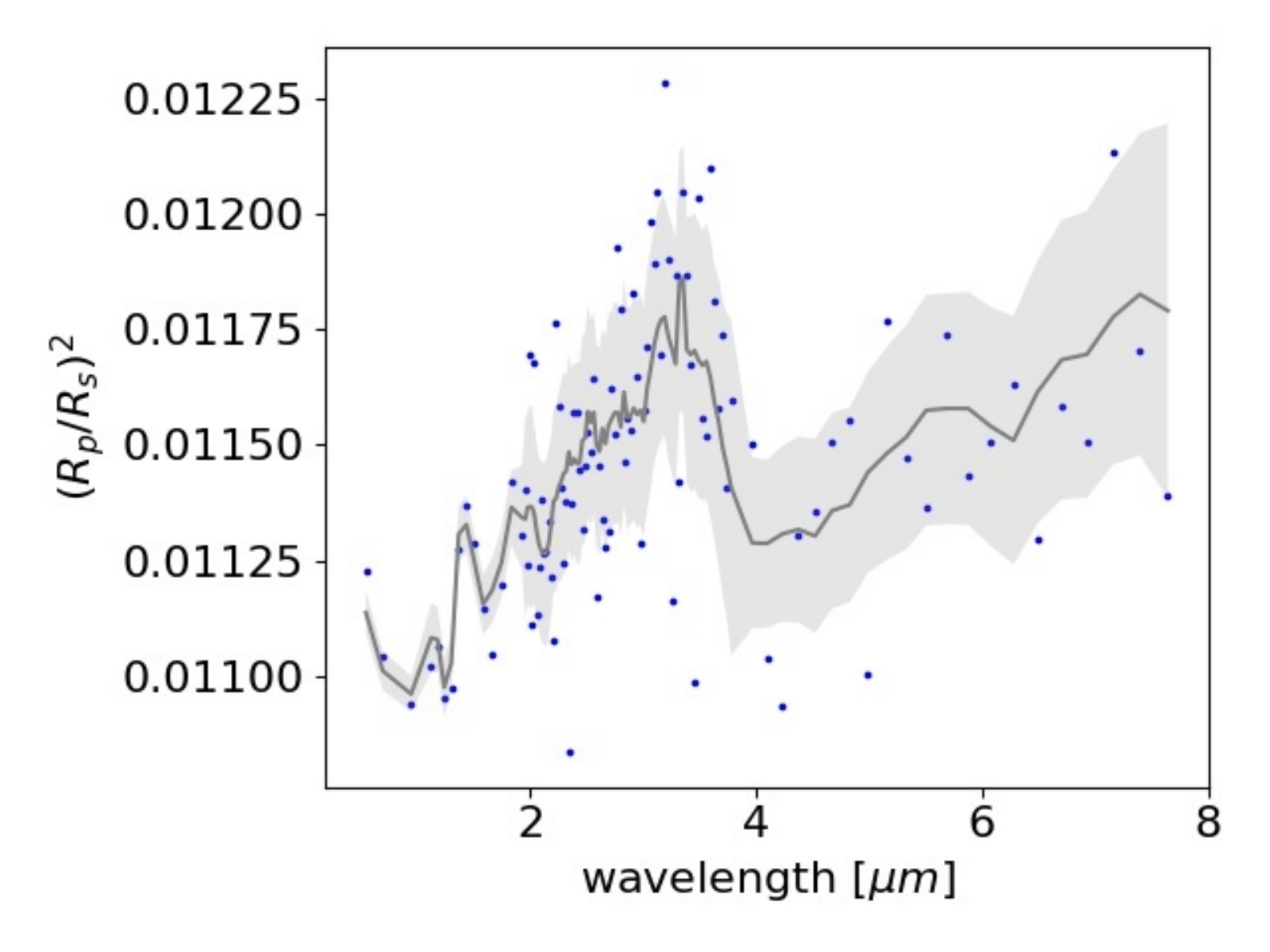}
		\caption{WASP 79b -like planet.}
	\end{subfigure}
	\caption{Example of simulated spectra. The grey solid lines are the noiseless spectra simulated and binned at \ARIEL\ Tier 3 spectral resolution. The grey bands are the $1-\sigma$  confidence levels centred around the simulated spectra for a number of transit observations needed to match the Tier 1 required SNR. The blue dots are noised data points representing  Tier 1 observed spectra. Starting from the left, the first planet is HD 209458 b-like, the second one is GJ 1214 b-like and the third one is WASP-79 b-like. Their atmospheres are built as described in Sec. \ref{sec:planetary population}.}
	\label{fig:example_spectra}
\end{figure}

Each planetary spectrum generated by Alfnoor is binned at \ARIEL's Tier 3 spectral resolution. These spectra make up the ``noiseless spectra'' data set. ArielRad then predicts the noise for each spectral bin at the Tier resolution. To reproduce a Tier 1 observation we scatter the data around the true value according to a normal distribution with the mean coinciding with the simulated spectrum, and a standard deviation equal to the noise estimated with ArielRad at each spectral bin. This noise is a re-scaled version of the Tier 3 noise, obtained by combining the number of transit observations needed to match the Tier 1 required SNR. Using these scattered spectra, we build the ``observed spectra'' data set. Examples of the resulting spectra are shown in Fig. \ref{fig:example_spectra}.

We generate POP-I using the full 1000 planets candidate list and we produce one realisation for each planet. A similar approach was used by  \citet{Changeat2020} in their investigation of the \ARIEL\ Tier-2 observations.
We use the POP-I population to test the strategies described later in the text.  

\paragraph{POP-II}
We produce another data set keeping the same 1000 planets from the target list and the randomisation rules of POP-I. However, this time we modify the chemical composition to include only H$_2$O and CH$_4$. We use POP-II to perform tests against a simpler population, as detailed later in the text.

\paragraph{POP-III}
To build the last population, we use the same list of 1000 planets, where each planet is repeated 4 times, such that there are 4 randomised atmospheres for each unique set of stellar and planetary properties that defines a planet. While the temperature and clouds conditions used are the same as those discussed for POP-I, for each molecule we widen the abundance boundaries to $10^{-9} \to 10^{-2}$ on a uniform logarithmic scale. We call this population POP-III, and we use it to train our machine learning algorithms. 

\subsection{Flat planet detection}
\label{sec:flat_spectra}

The first goal of this work, as listed in Sec. \ref{sec:intro}, is to identify featureless spectra. This will help in the  selection of targets to be re-observed in \ARIEL's higher Tiers.
Given the property of the \ARIEL\ payload, we divide the spectral wavelength range in four parts or bands: 
\begin{itemize}
	\item from $0.5$ to $1.1 \; \mu m$, sampled by three photometers;
	\item from $1.1$ to $1.95 \; \mu m$, corresponding to the NIRSpec wavelength range;
	\item from $1.95$ to $3.9 \; \mu m$, corresponding to the AIRS-CH0 wavelength range;
	\item from $3.9$ to $7.8 \; \mu m$, corresponding to the AIRS-CH1 wavelength range.
\end{itemize}

For every planet, and for every band we estimate a $\chi^2$ using all measurements in the band to assess the compatibility with a flat, zero-gradient line: for each planet there are four $\chi^2$ estimates, one for each band above.  
We reject the hypothesis of spectral flatness in a given band with a $3-\sigma$ confidence if $\chi^2 > 1+3 \sqrt{\frac{2}{\nu}}$, where $\nu$ are the degrees of freedom. Therefore, if any of the four bands has a $\chi^2$ smaller than this number, we mark the band as flat. If a planetary spectrum has all 4 bands marked as flat, it is classified as a flat spectrum. This strategy is similar to that presented in \cite{Zellem2019}, however, while in that work the authors were only focused on the \ARIEL\ FGS channels, here we are considering the full \ARIEL\ spectral coverage.

\subsection{An optimised molecular metric}
\label{sec:metric}

\begin{figure}
	\centering
	\includegraphics[width = 0.5\textwidth]{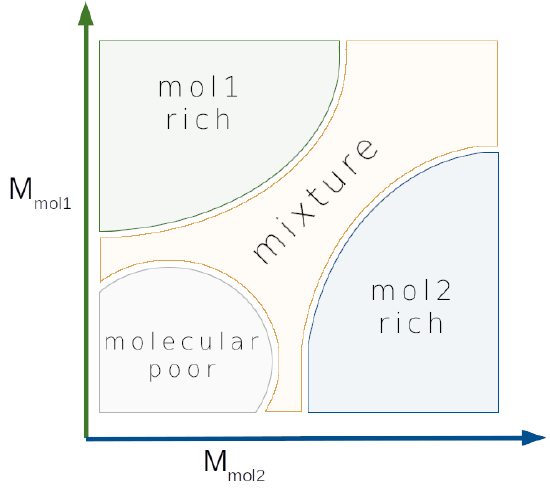}
	\caption{Illustration of the diagram we expect to build with our metric. Here, the metric is used to compare two molecules, $mol1$ and $mol2$. By drawing $M_{mol1}$ versus $M_ {mol2}$, we aim to separate four different regions: one rich in the first molecule at the top left (green), where $M_{mol1}$ grows and $M_{mol2}$ is low: a similar region at the bottom right (blue), where the planet atmosphere is rich in the second molecule, because $M_{mol2}$ is high and $M_{mol1}$ is low; a region where molecular poor planets are located (grey), or those that have no features in the considered bands, where both $M_{mol1}$ and $M_{mol2}$ are low; a region for mixed atmosphere (yellow) in the central portion of the diagram.}
	\label{fig:expected_diagram}
\end{figure}

The second goal listed in Sec. \ref{sec:intro} is to develop a metric, $M_{mol}$, to assess the presence of a molecule, $mol$, in the planets atmosphere. We want this metric to work in such a way that by comparing two molecules, the metric produces a diagram similar to that in Fig. \ref{fig:expected_diagram}. In the diagram we can distinguish four regions: two regions where the atmospheres are rich in a single molecule and therefore only show its characteristic features; a third region where the atmospheres show features from both molecules; a fourth region where features are absent, either because the planets have flat spectra or because the features from both molecules do not emerge from a thick layer of clouds. 

To compare different planets and constrain their atmospheric molecular content, the metric should be (i) sensitive to the spectral signature of molecules, (ii) independent of the planet size, and (iii) independent of the scale height. Here we present a metric that fulfils these 3 conditions and we show its current limitations. 

For each molecule, we select $N$ bands within the \ARIEL\ wavelength range, where the molecular features in the transmission spectrum are strong. Then, for each planet, we compute the average in each band, $S_{band_i}$ and its dispersion, $\sigma_{band_i}$. 

\begin{equation} \label{eq:sband}
S_{band_i} = \frac{1}{M}\sum_j^M S_j
\end{equation}
\begin{equation} \label{eq:sigma_sband}
\sigma_{band_i} = \sqrt{\frac{1}{M} \sum_j^M (S_j - S_{band_i})^2}
\end{equation}
where $M$ is the number of spectral bins in the band, $S_j$ is the atmospheric transmission spectrum estimated in the $j^{\rm th}$ wavelength bin. 

We do the same with a control band where we know there are no major molecular features from the molecule considered, called ``normalisation band'', obtaining $S_{norm}$ and $\sigma_{norm}$. 
We select a different normalisation band for each molecule (Tab. \ref{tab:feature_bands}).

Thus, for each molecule, $mol$, we define

\begin{equation} \label{eq:mmol}
M_{mol} = \frac{1}{N} \sum_i^N \frac{S_{band_i}-S_{norm}}{\sqrt{\sigma_{band_i}^2+\sigma_{norm}^2}}
\end{equation}
Defined in this way, $M_{mol}$ is similar to a signal-to-noise ratio, where the signals are the molecular features arising above the ``normalisation band'', and the noise is the dispersion in the band. Therefore,



\begin{equation} \label{eq:sigma_mmol}
\sigma_{M_{mol}} = \frac{1}{\sqrt{N}}
\end{equation}

The metric thus designed, by averaging the contribution of $N$ different bands, corresponding to $N$ different features of the same molecule,  reduces the chance to be misled by overlapping features in one of the bands considered. As \ARIEL's Tier 1 is optimised for low resolution spectroscopy, spectral binning increases the SNR. 
Also, this metric is (i) sensitive to the presence of molecules, (ii) independent of the planet size, and (iii) independent of the scale height (see Appendix \ref{sec:analytic_metric} for details),  at the cost of the introduction of a bias: eq. \ref{eq:sigma_sband} provides an estimate of the spectral dispersion when applied to noiseless spectra, and it is larger for observed spectra because of the presence of measurement noise. 
Therefore, the absolute value of $M_{mol}$ of eq. \ref{eq:mmol} is always smaller on observed spectra compared to noiseless spectra of the same planet.  While the bias effects are further discussed in Sec. \ref{sec:first_outcomes}, we note here that a detailed characterisation of the instrumental noise would allow to de-bias the metric, but we leave this investigation to future work, and we focus the attention on the performance of the metric in extracting information from Tier 1 observations. 

To maximise the metric efficiency, the challenge is to identify the best performing wavelength range to use: large enough to reduce the uncertainty introduced by the observational noise, but small enough to distinguish the molecular features of interest. 

\begin{figure}
	\centering
	\begin{subfigure}[b]{0.32\textwidth}
		\centering
		\includegraphics[width=\textwidth]{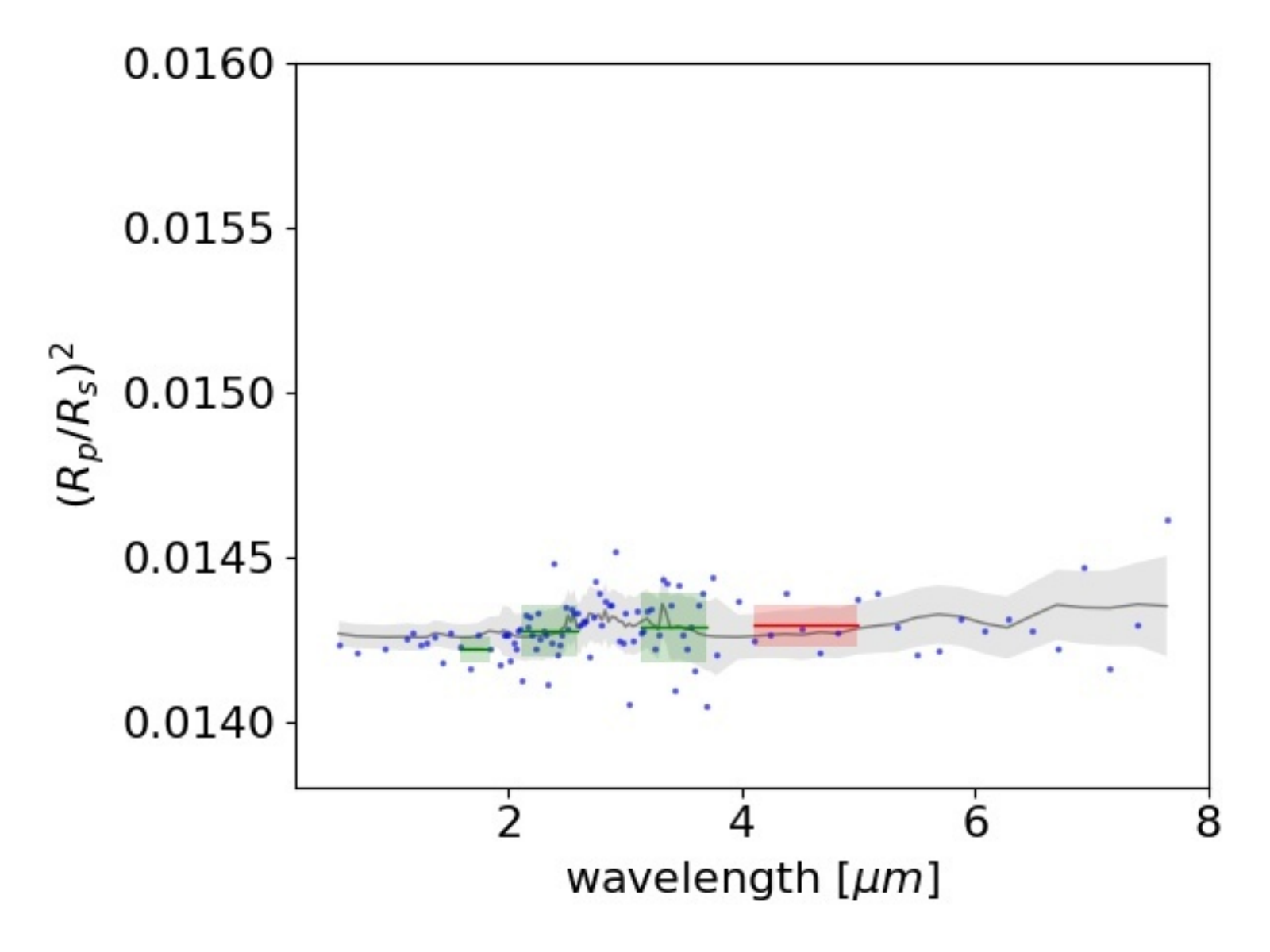}
		\caption{Overcast HD 209458b -like planet with $M_{CH_4}$ data bands highlighted.}
	\end{subfigure}
	\hfill
	\begin{subfigure}[b]{0.32\textwidth}
		\centering
		\includegraphics[width=\textwidth]{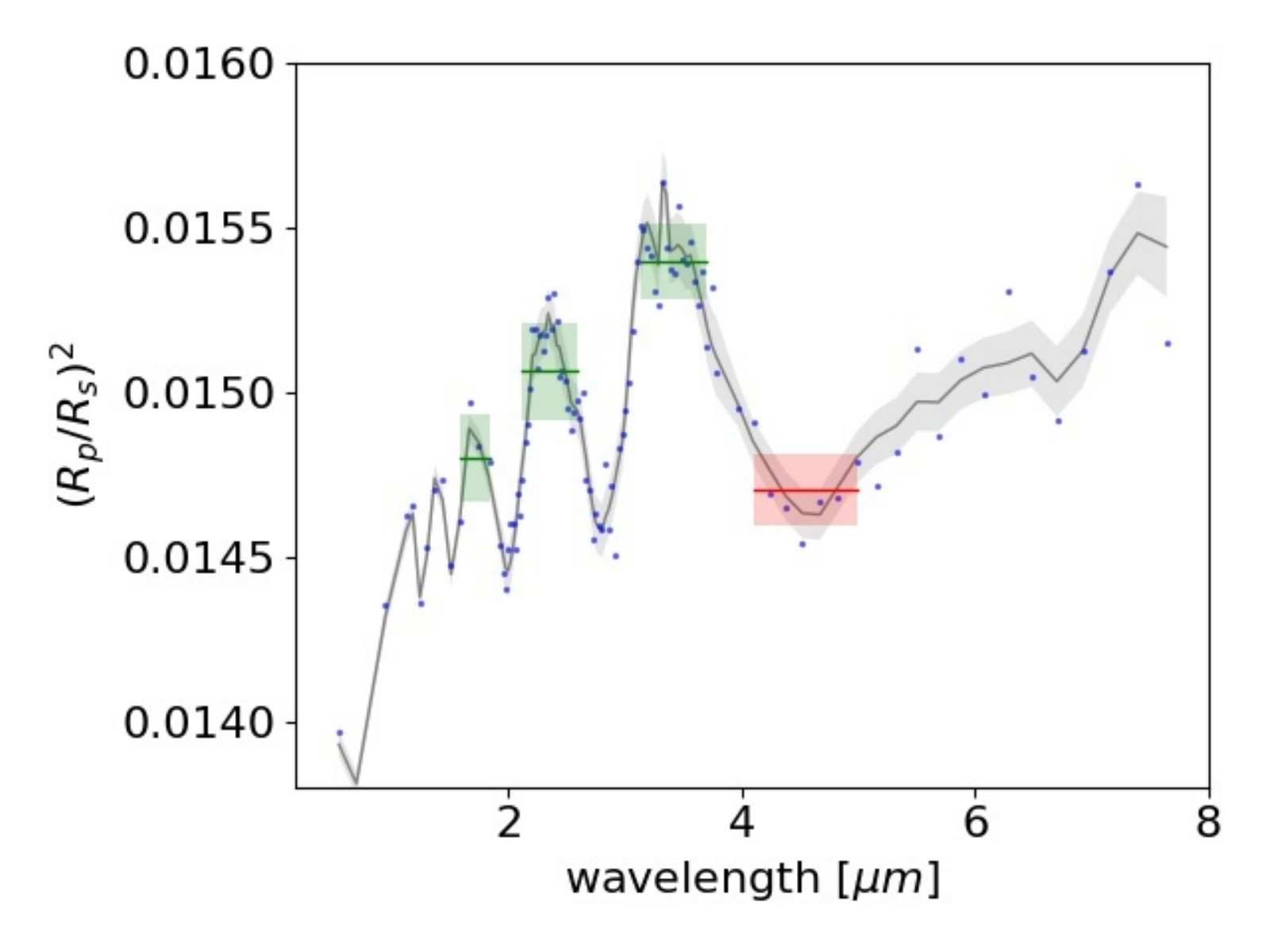}
		\caption{Methane rich HD 209458b -like planet with $M_{CH_4}$ data bands highlighted.}
	\end{subfigure}
	\hfill
	\begin{subfigure}[b]{0.32\textwidth}
		\centering
		\includegraphics[width=\textwidth]{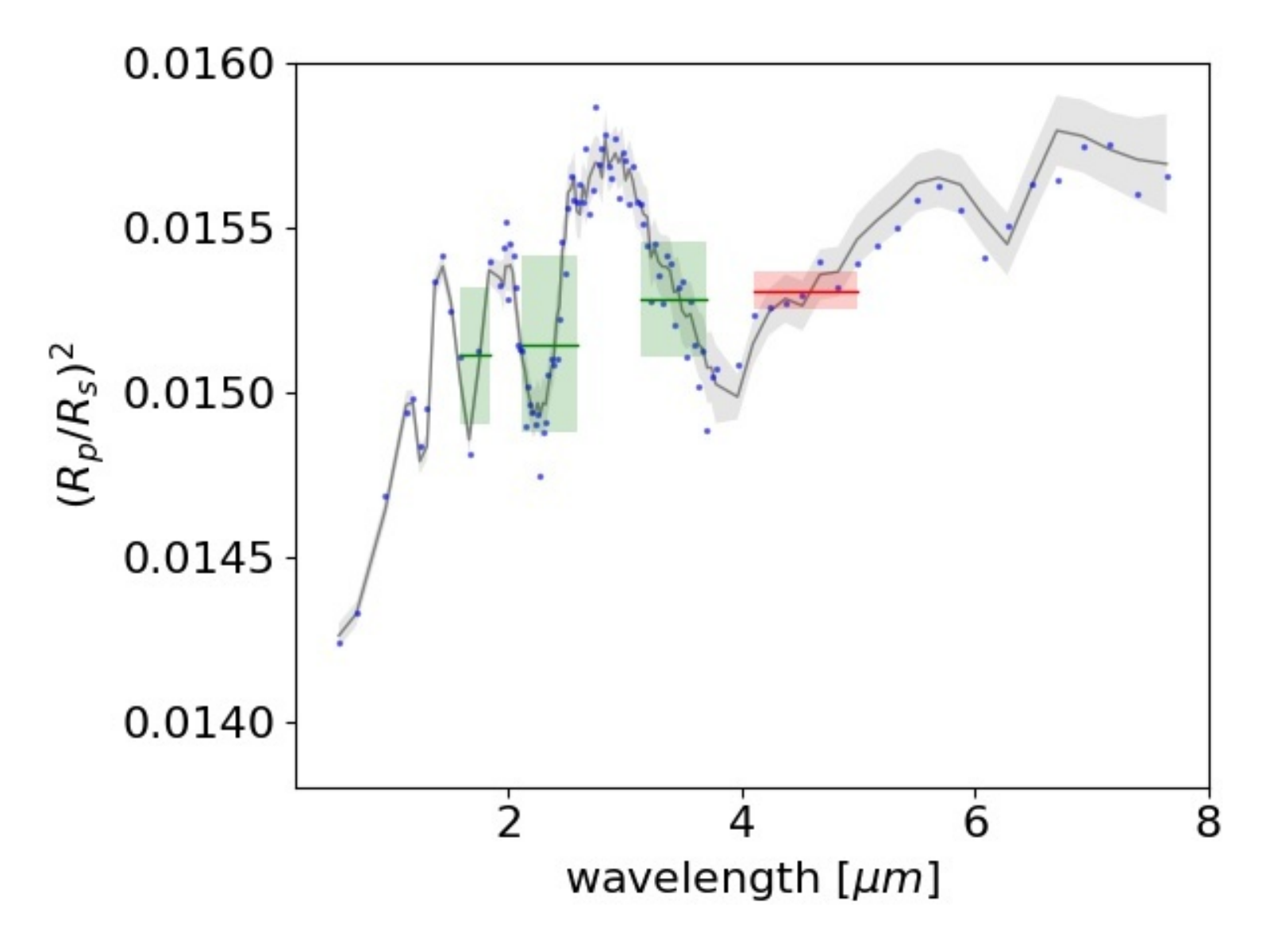}
		\caption{Water rich HD 209458b -like planet with $M_{CH_4}$ data bands highlighted.}
	\end{subfigure}
	\hfill
	\begin{subfigure}[b]{0.32\textwidth}
		\centering
		\includegraphics[width=\textwidth]{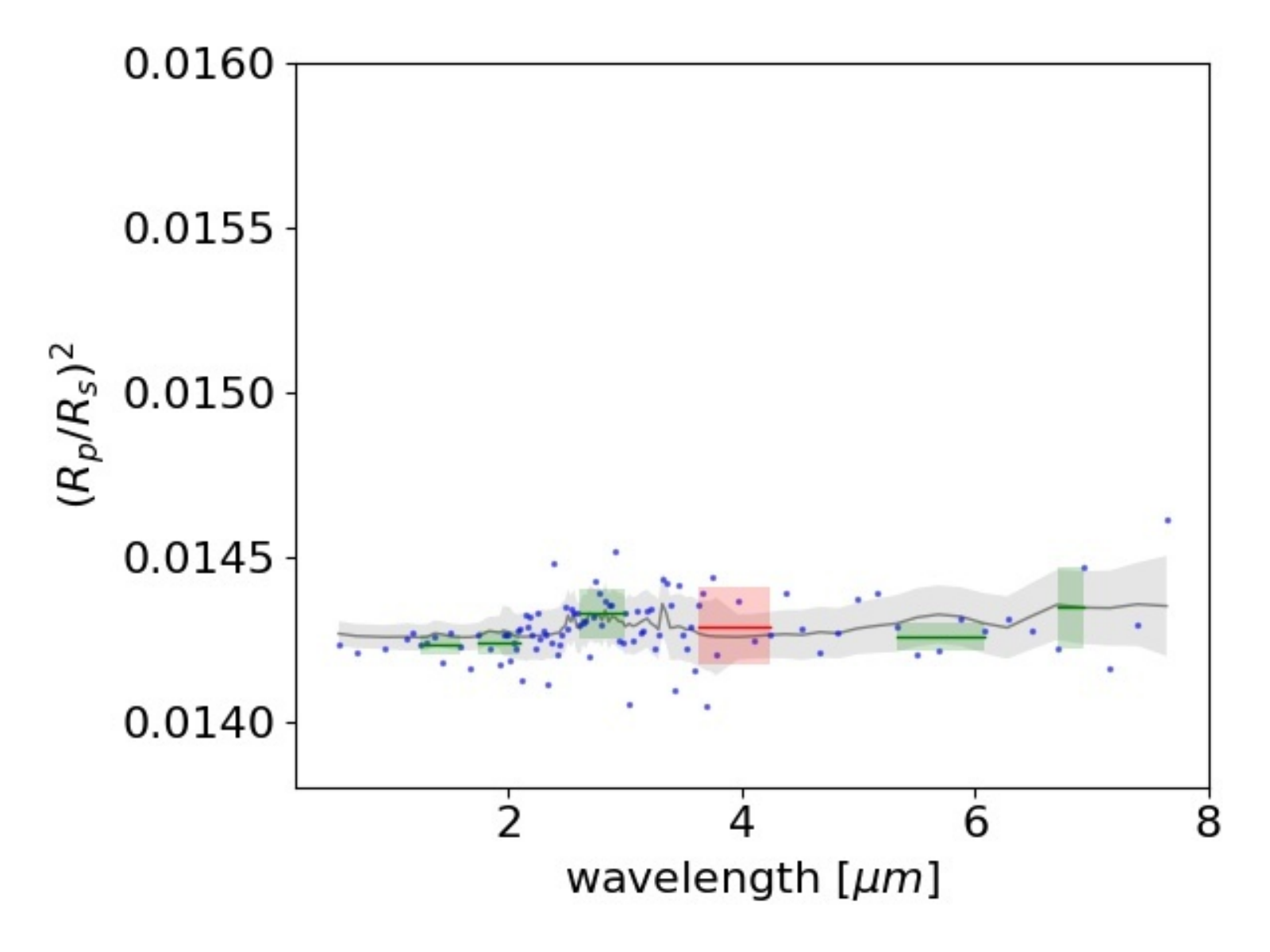}
		\caption{Overcast HD 209458b -like planet with $M_{H_2O}$ data bands highlighted.}
	\end{subfigure}
	\hfill
	\begin{subfigure}[b]{0.32\textwidth}
		\centering
		\includegraphics[width=\textwidth]{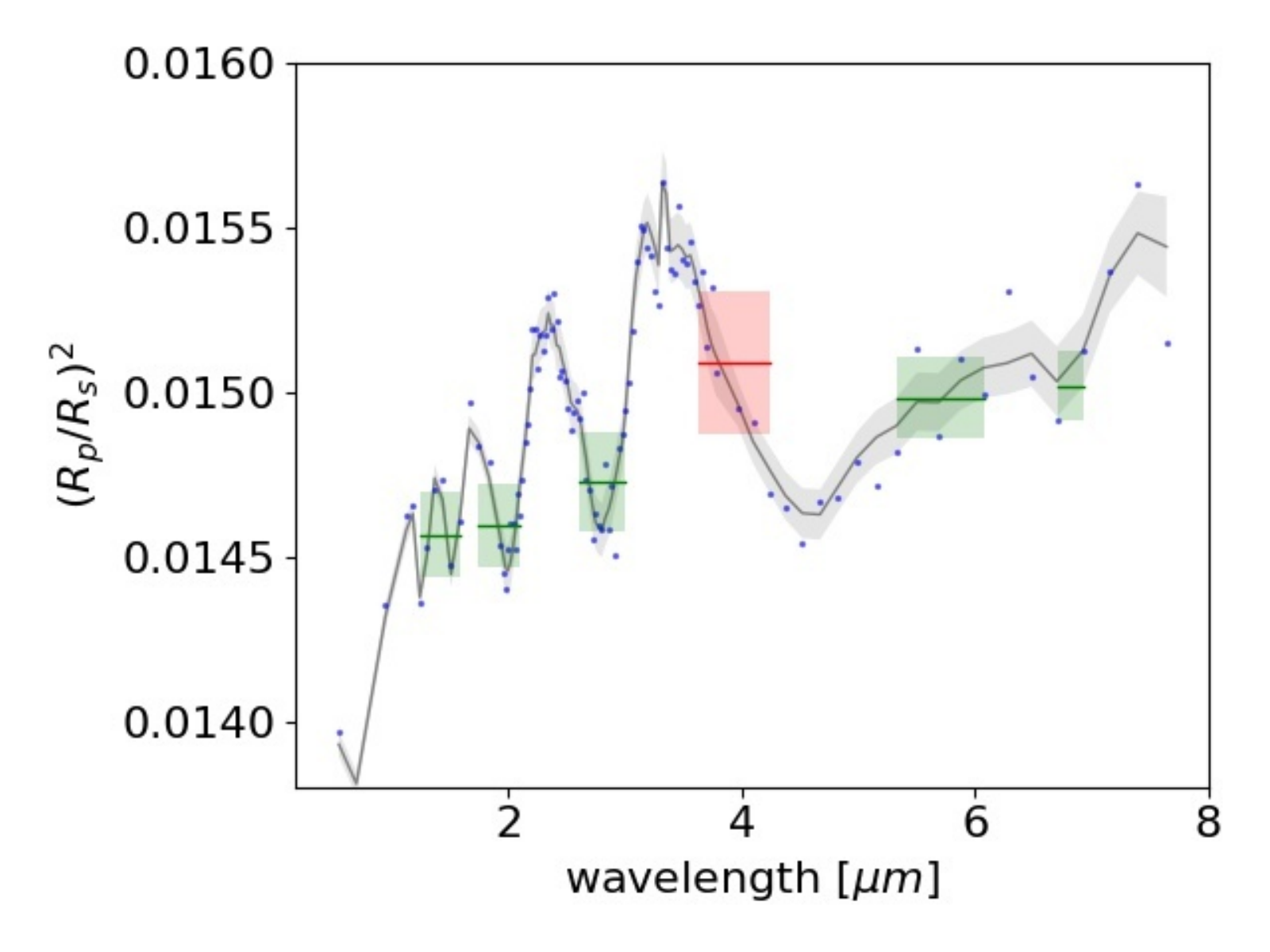}
		\caption{Methane rich HD 209458b -like planet with $M_{H_2O}$ data bands highlighted.}
	\end{subfigure}
	\hfill
	\begin{subfigure}[b]{0.32\textwidth}
		\centering
		\includegraphics[width=\textwidth]{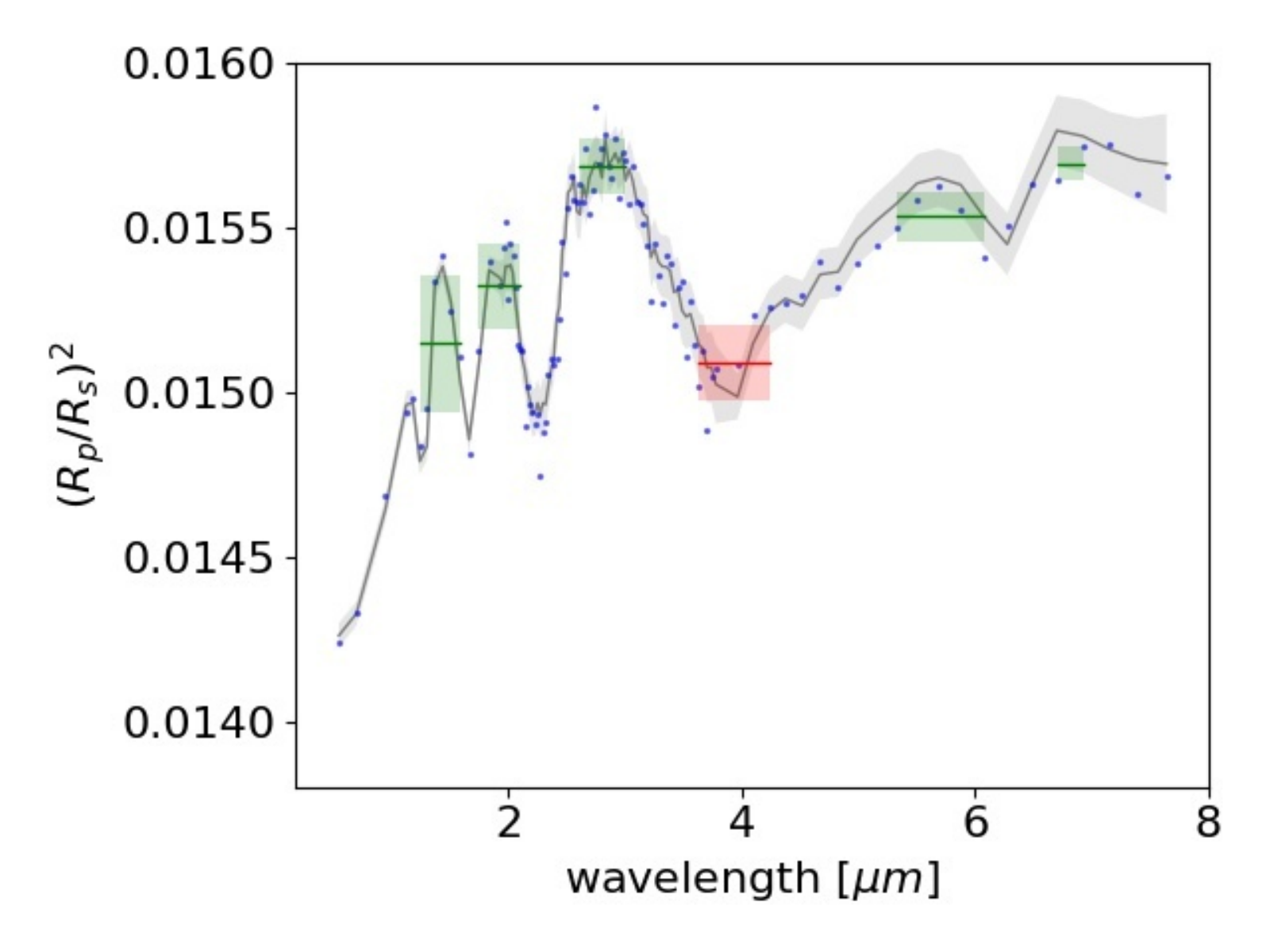}
		\caption{Water rich HD 209458b -like planet with $M_{H_2O}$ data bands highlighted.}
	\end{subfigure}
	
	\caption{Here are shown three examples of randomised spectra. For the same planet, HD 209458 b, we present three different realisations: a flat atmosphere (first column), a methane rich atmosphere (second column) and a water rich atmosphere (third column). Each column shows the same planetary spectra. Grey solid lines are the original binned spectral data (Tier 3 spectral resolution), the filled grey areas are the $1-\sigma$ uncertainties (Tier 1), and blue dots are the simulated observation data used in this work. The top row highlights the $M_{CH_4}$ feature bands from Tab. \ref{tab:feature_bands}, while the bottom row shows the $M_{H_2O}$ bands. In green are reported the molecular feature bands values, with their dispersion, while in red are reported the normalisation bands'. Comparing the rows we see how the bands selected match the relevant molecular spectral features.}
	\label{fig:example_spectra_HD}
\end{figure}

In this work, we consider only H$_2$O, CH$_4$ and CO$_2$, and the bands used are listed in Tab. \ref{tab:feature_bands}. Even though NH$_3$ is present in our sample, it is used only to introduce a nuisance and challenge our metric, because  NH$_3$ has features overlapping with those of water.  We use 3 feature bands for CH$_4$ and CO$_2$ and 5 for H$_2$O.
Examples of the bands used for CH$_4$ and H$_2$O are shown in Fig. \ref{fig:example_spectra_HD} where, for the same planetary template, HD 209458 b, we simulate different atmospheres (overcast, CH$_4$ rich and H$_2$O rich) to show how the metric captures the relevant spectroscopic features.

In the next section, we show how we intend to use this metric to build a diagram similar to that of Fig. \ref{fig:expected_diagram}.

\begin{table}[]
	\caption{Wavelength ranges used to select the molecular features in the spectra (left table) and the normalisation bands (right table) for H$_2$O, CH$_4$ and CO$_2$.}
	\label{tab:feature_bands}
	\begin{subtable}{0.50\textwidth}{
			
			\centering
			\begin{tabular}{lll}
				\hline\noalign{\smallskip}
				{\bf \centering H$_2$O} & {\bf \centering CH$_4$} & {\bf \centering CO$_2$} \\ 
				\noalign{\smallskip}\hline\noalign{\smallskip}
				$1.2 \to 1.6 \; \mu m$ & $1.5 \to 1.8 \; \mu m$ & $1.9 \to 2.3 \; \mu m$ \\
				$1.7 \to 2.1 \; \mu m$ & $2.1 \to 2.6 \; \mu m$ & $2.6 \to 3.2 \; \mu m$  \\
				$2.6 \to 3.0 \; \mu m$ & $3.1 \to 3.7 \; \mu m$ & $4.2 \to 4.8 \; \mu m$  \\
				$5.4 \to 6.1 \; \mu m$ & & \\
				$6.5 \to 7 \; \mu m$ & & \\
				\noalign{\smallskip}\hline
			\end{tabular}
		}
	\end{subtable}
	\hspace{1.2 cm}
	\begin{subtable}{0.2\textwidth}
		{
			\centering
			\begin{tabular}{ll}
				\hline\noalign{\smallskip}
				{\bf \centering Molecule} & {\bf \centering Normalisation}\\ 
				\noalign{\smallskip}\hline\noalign{\smallskip}
				H$_2$O & $3.6 \to 4.2 \; \mu m$ \\
				CH$_4$ & $4.0 \to 5.0 \; \mu m$ \\
				CO$_2$ & $5.0 \to 6.0 \; \mu m$ \\
				\noalign{\smallskip}\hline				
				
			\end{tabular}
		}
	\end{subtable}

\end{table}

\subsubsection{Planets classification \label{sec:KNN}}

The metric requires to be calibrated to assess its capability to estimate the presence of a molecule. The final product is a diagram similar to Fig. \ref{fig:expected_diagram}, that can be used as a look-up table, such that, given an observed spectrum, its corresponding  $M_{mol}$ can be located on the diagram, and its possible composition inferred. 


To assess the ability of the metric to separate the atmospheres in the sample, we use the \textit{k}-nearest neighbours (KNN) algorithm, a non-parametric pattern recognition algorithm \citep{Hastie2009}. This algorithm, after a training process, assigns a class to an element given the properties of its neighbours. The goal is to classify observed spectra by their molecular content, according to their $M_{mol}$. Considering two molecules at a time, we first define four classes of planets: molecular poor, $mol1$ rich, $mol2$ rich and mixture, as defined in  Tab. \ref{tab:classes}.

\begin{table}[]
	\centering
	\caption{Diagram classes and conditions.}
	\label{tab:classes}
	
	\begin{tabular}{ll}
		\hline\noalign{\smallskip}
		{\bf \centering Class} & {\bf \centering Condition}\\ 
		\noalign{\smallskip}\hline\noalign{\smallskip}
		molecular poor &  $Ab_{mol1}<10^{-5}$ and $Ab_{mol2}<10^{-5}$\\
		$mol1$ rich & $Ab_{mol1}>10^{-4}$ and  $Ab_{mol1}>10 \times Ab_{mol2}$\\
		$mol2$ rich &  $Ab_{mol2}>10^{-4}$ and $Ab_{mol2}>10 \times Ab_{mol1}$\\
		mixture & everything else \\
		\noalign{\smallskip}\hline
	\end{tabular}
\end{table}

The KNN algorithm used classifies each planet according to the 20 (k = 20) nearest planets, in the $M_{mol1}$ vs $M_{mol2}$ space, in the same data set. We choose to use 20 neighbours ($2\%$ of the full data set) to minimise the number of misclassified planets. The closest neighbours are uniformly weighted, and we verified that weighting the neighbours with their Euclidean distance in the metric space does not affect the results significantly.

\begin{figure}
	\centering
	\includegraphics[width = 0.99\textwidth]{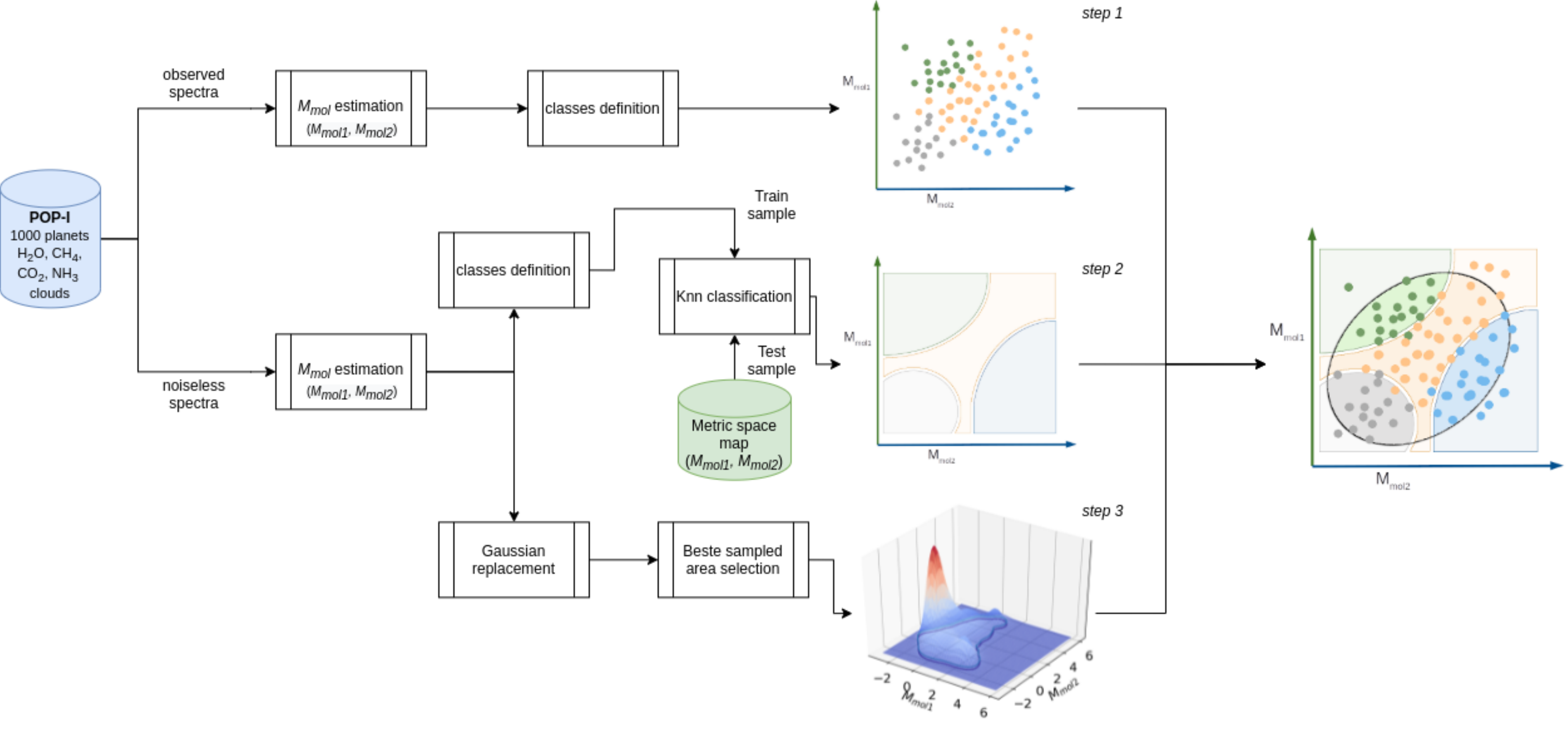}
	\caption{Planets classification summary. The figure reports the steps implemented to build the diagram in Fig. \ref{fig:expected_diagram}. Starting from POP-I, for each planet we compute $(M_{mol1}, M_{mol2})$ for the considered molecules and for both observed and noiseless data. 
    Following the top branch, classes are assigned to the observed spectra (\textit{step 1} in the text). 
    Following the middle branch, a KNN classification is performed on noiseless spectra to calibrate the metric space (\textit{step 2} in the text).
    Following the bottom branch, the distribution of noiseless metric data points is convolved with a 2D Gaussian with varying widths to generate a unit-normalised volume. The intersection between this volume and the calibration of  \textit{step 2} selects the best sampled (i.e. calibrated) region in the metric space (\textit{step 3} in the text).
    The combination of these three steps is shown in the rightmost diagram to be compared with Fig. \ref{fig:expected_diagram}.
    }
    
    
	\label{fig:knn_summary}
\end{figure}

The analysis involves three separated steps,  summarised in Fig. \ref{fig:knn_summary}, applied to POP-I.

\textit{Step 1.} We estimate the $(M_{mol1}, M_{mol2})$ on the POP-I observed spectra. We assign a class to each POP-I planet using its input molecular abundance values, $Ab_{mol}$, that are stored during the population production. This process is described in the top branch of Fig. \ref{fig:knn_summary}


\textit{Step 2.} 
To calibrate the metric, we map the metric space grid by training the KNN algorithm on the $(M_{mol1}, M_{mol2})$  estimated from the noiseless POP-I planetary spectra. We assign again a class to each planet using its input molecular abundance, $Ab_{mol}$, and the training is performed on a randomly chosen selection accounting for $70\%$ of the data set, while we use the remaining $30\%$ to test the success of the training. 
Finally, we classify each point $(M_{mol1}, M_{mol2})$ of the $M_{mol}$ space grid $M_{mol}$ sampled at a step width of $0.2 \, M_{mol}$, obtaining a map comparable to Fig. \ref{fig:expected_diagram}. This part of the procedure corresponds to the central branch of Fig. \ref{fig:knn_summary}.

\textit{Step 3.} Since the noiseless planetary spectra are not expected to sample the parameter space uniformly, 
we build a mask to select a region of the $(M_{mol1}, M_{mol2})$ space that is sufficiently well sampled to achieve a reliable classification.
To do so, we replace each $(M_{mol1}, M_{mol2})$ point representing a noiseless planetary spectrum with a two-dimensional Gaussian distribution using the metric dispersion in the two directions as $\sigma$. We sum the Gaussian volumes on the parameter space, ending up, after volume normalisation, with a statistical distribution of our data points on the parameter space grid. Then, we select a region in the metric space that results in a total volume of $95\%$, therefore removing all under-sampled areas from the grid. This last step is represented in the bottom branch of Fig. \ref{fig:knn_summary}.

The combination of the three steps is shown in the rightmost panel of Fig. \ref{fig:knn_summary} and it is the equivalent of Fig. \ref{fig:expected_diagram} calibrated for the metric on the investigated population.

\subsection{Deep and Machine Learning}
\label{sec:deep_learning}
The metric presented in Sec. \ref{sec:metric} is based on binning the spectra, and therefore is equivalent to using \ARIEL\ as a multi-band photometer. This strategy is in line with the Tier 1 definition of \citet{Tinetti2018}. However, we are also investigating different strategies to classify spectra by their molecular content (third goal listed in Sec. \ref{sec:intro}).  
Deep Learning and Machine Learning (ML) techniques are promising because these algorithms can learn to classify planets from their spectral shape over the whole wavelength range sampled by \ARIEL. 
Another advantage over the metric is that ML techniques are not supposed to be biased by the instrumental noise, or at least they can be made to learn how to deal with the bias provided that a sufficiently large and representative set of examples is provided in training. 
To train the algorithms we use the POP-III observed spectra and their known abundances as a training sample. Each example spectrum is normalised to zero mean and unit dispersion. The normalisation facilitates the training process but might introduce a bias that may be very similar to that affecting the metric. A detailed investigation of these aspects concerning ML is left to future work. Knowing the input abundance of each planet, $Ab_{mol}$, we can define a threshold and flag a planet as bearing a certain molecule if $Ab_{mol}$ is larger than the threshold. This means that, for each molecule, the algorithm learns to flag the planets as bearing that molecule by looking at characteristic spectral shapes. Then we measure the algorithm ability to ``learn'' by how much they can generalise their predictions to unknown shapes, testing it on POP-I observed spectra, used as a test data set. The comparison of the ML classification with the known input abundance of each POP-I planet provides an estimate of the success rate. 

A detailed investigation of the use of these algorithms and their limitations will be discussed in future work: here we report only an example of how these tools might be used and we compare some preliminary results with the outcomes of the metric of Sec. \ref{sec:metric}. We implemented all algorithms in Python using the scikit-learn\footnote{\url{https://scikit-learn.org/0.22/}} package presented in \citet{scikit-learn}.

The first ML algorithm we use is the KNN algorithm described above. This time we want to simply classify the planets and not to produce a map as in Sec. \ref{sec:KNN}. For this exercise, we use the scikit-learn default KNN setting: $k=5$ and uniform weight for the neighbours. 
Other Machine Learning algorithms can be used to classify planets. Here we also present our preliminary results using a Multi-layer Perceptron (MLP) classifier, a Random Forest Classifier (RFC) and a Support Vector Classifier (SVC) \citep[e.g.][]{Goodfellow2016,Sturrock2019}.
The MLP is a feed-forward neural network composed of multiple layers of perceptrons largely used in classification problems. To produce the results shown later in the text we use an MLP network keeping the scikit-learn default settings (a single hidden layer made of 100 units) and we classify the spectra with the same procedure used for the KNN.
The RFC is an ensemble of decision trees used for classification, where each decision tree is a directed graph and each vertex is a binary test. In this work, we use an RFC set-up commonly used in binary decision problems, which has a number of features equal to the square root of the number of input data points, again, as per scikit-learn is the default configuration.
The SVC is a Support Vector Machine method, a family of non-probabilistic linear classifiers that construct hyper planes to separate the data points. For the aim of this paper, we implemented a simple SVC shaping the decision function in ``one-vs-one'' mode, as it is the default configuration in scikit-learn at the moment of writing.

\section{Results}
\label{sec:results}

\subsection{Identifying flat spectra}
\label{sec:flat_results}

\begin{figure}
    \centering
    \includegraphics[width=0.8\textwidth]{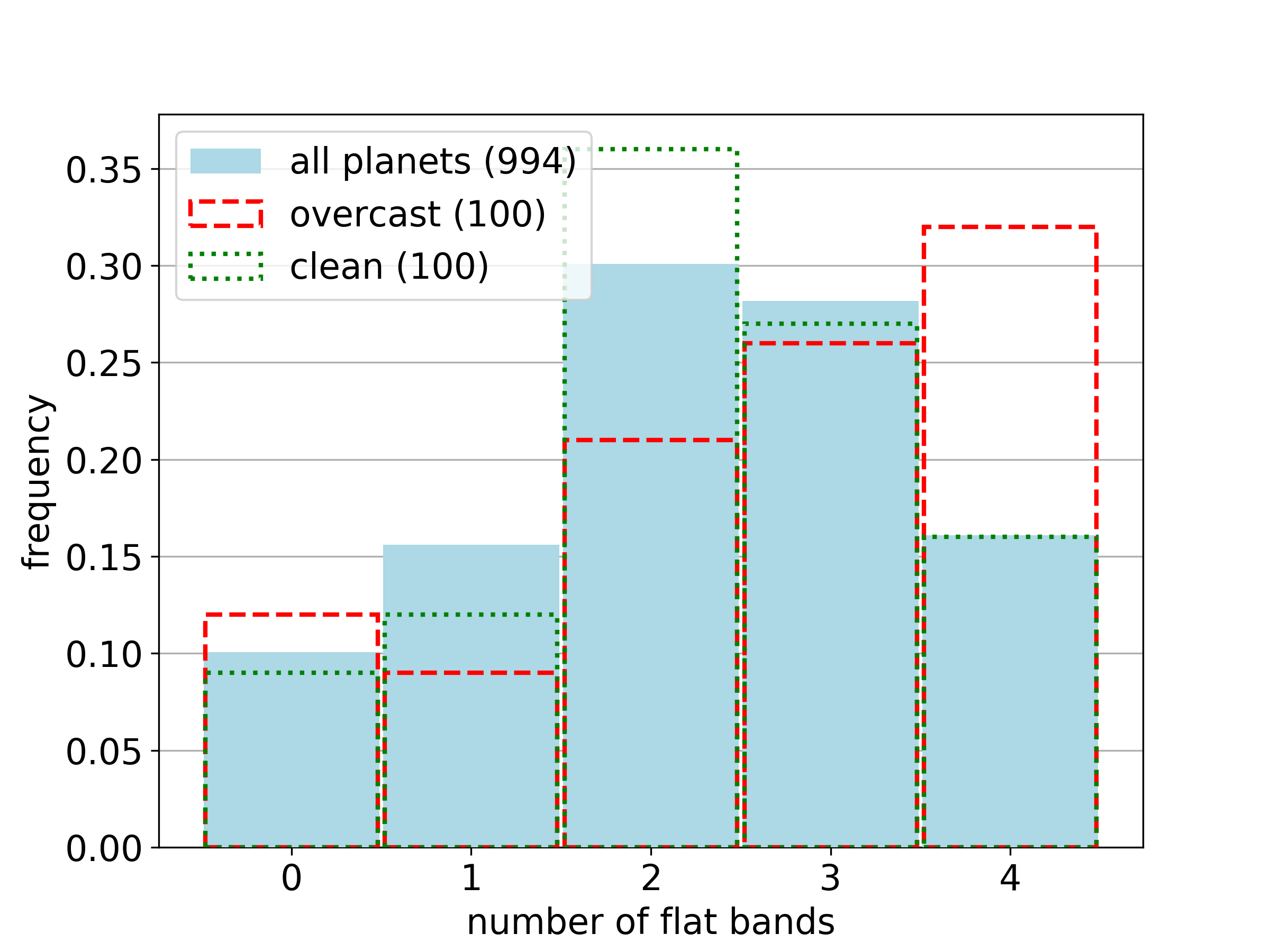}
    \caption{The histograms show the frequency of planets in the population vs the number of flat bands. We consider four bands: one for the photometers (VisPhot, FGS1, FGS2) and one for each spectrometer (NIRSpec, AIRS CH0 and AIRS CH1). Each band is compared with a constant value using a $\chi^2$ test to determine its compatibility with flatness. The light blue histogram shows the frequency of planets in the POP-I population with flat bands. The red dashed histogram shows the same statistic but for a selection of the 100 planets of POP-I that are more overcast. The green dotted histogram shows the opposite situation, for a selection of the 100 planets in POP-I for which the cloud pressure surface is the lowest (see text for details). We notice that the overcast planets show more flat bands than planets with fewer clouds.}
    \label{fig:flat_stats}
\end{figure}

Shown in Fig. \ref{fig:flat_stats} is the frequency of observed planets in the POP-I population that have a certain number of flat bands. In this population, $16 \%$ planets are to be considered ``flat'' as all of the four spectral bands considered are flat. From the figure, we notice that around $46\%$ of the planets in the population have three or more flat bands, which is consistent with POP-I known properties and with the ground truth \citep{Tsiaras2018,Iyer2016}, as mentioned in Sec. \ref{sec:planetary population}. 
In the same figure it is shown the same statistic for the 100 planets of POP-I most covered in clouds (corresponding to a cloud surface pressure of roughly $<10^{3}$ Pa), and for the 100 planets of POP-I with fewer clouds (corresponding to a cloud surface pressure of roughly $>10^{5.5}$ Pa). This comparison shows how overcast planets averagely present more flat bands than clean planets, demonstrating how this approach is sensitive to the presence of clouds.

This result clearly shows that Tier 1 observations are effective in the identification of atmospheres with no detectable molecular absorption features. 

\subsection{Spectra classification}
\label{sec:first_results}

\begin{figure}
	\centering
	
	\begin{subfigure}[b]{0.48\textwidth}
		\centering
		\includegraphics[width=\textwidth]{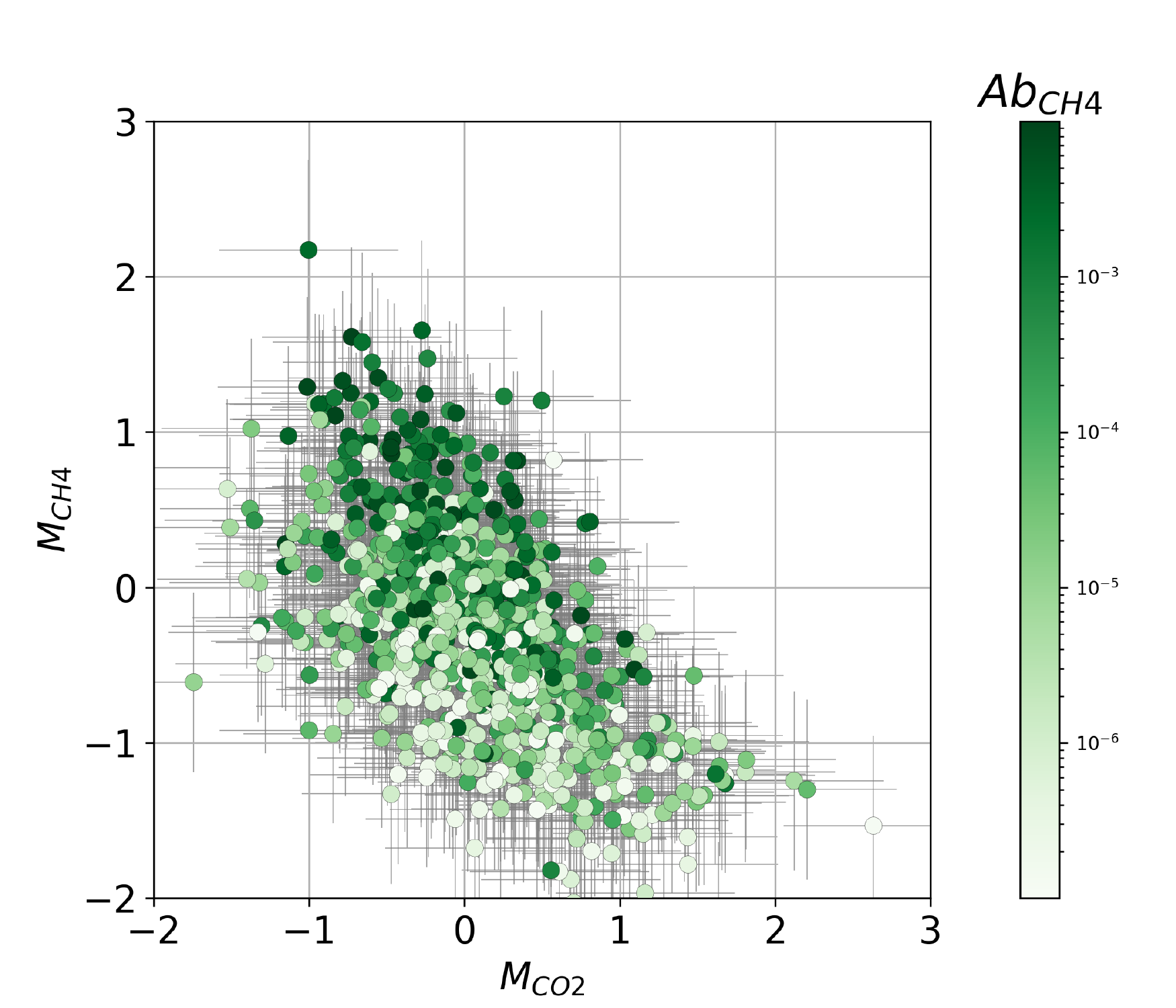}
		\caption{POP-I estimated $M_{CO_2}-M_{CH_4}$ coloured by CH$_4$.}
	\end{subfigure}
	\hfill
	\begin{subfigure}[b]{0.48\textwidth}
		\centering
		\includegraphics[width=\textwidth]{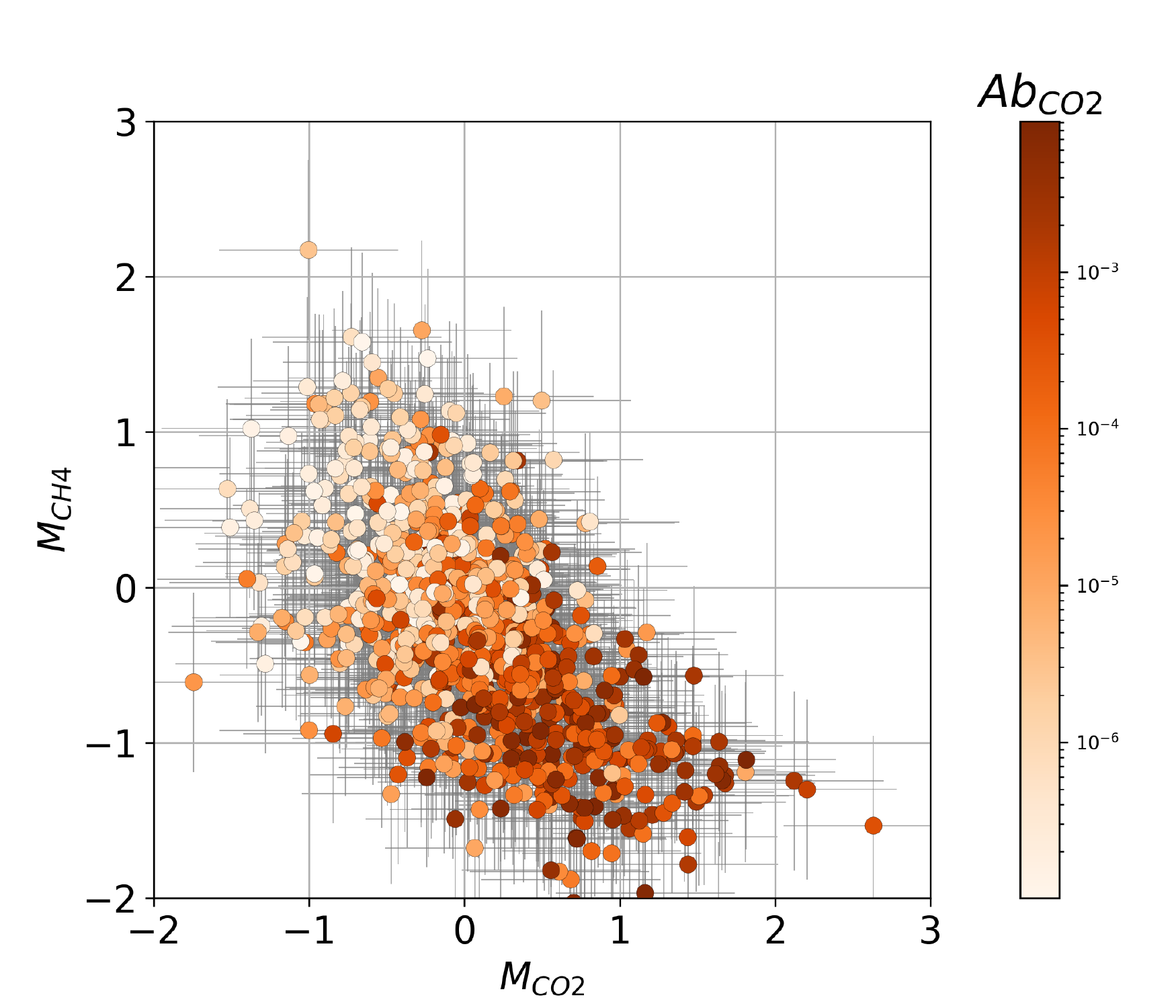}
		\caption{POP-I estimated $M_{CO_2}-M_{CH_4}$ coloured by CO$_2$.}
	\end{subfigure}  
	\begin{subfigure}[b]{0.48\textwidth}
		\centering
		\includegraphics[width=\textwidth]{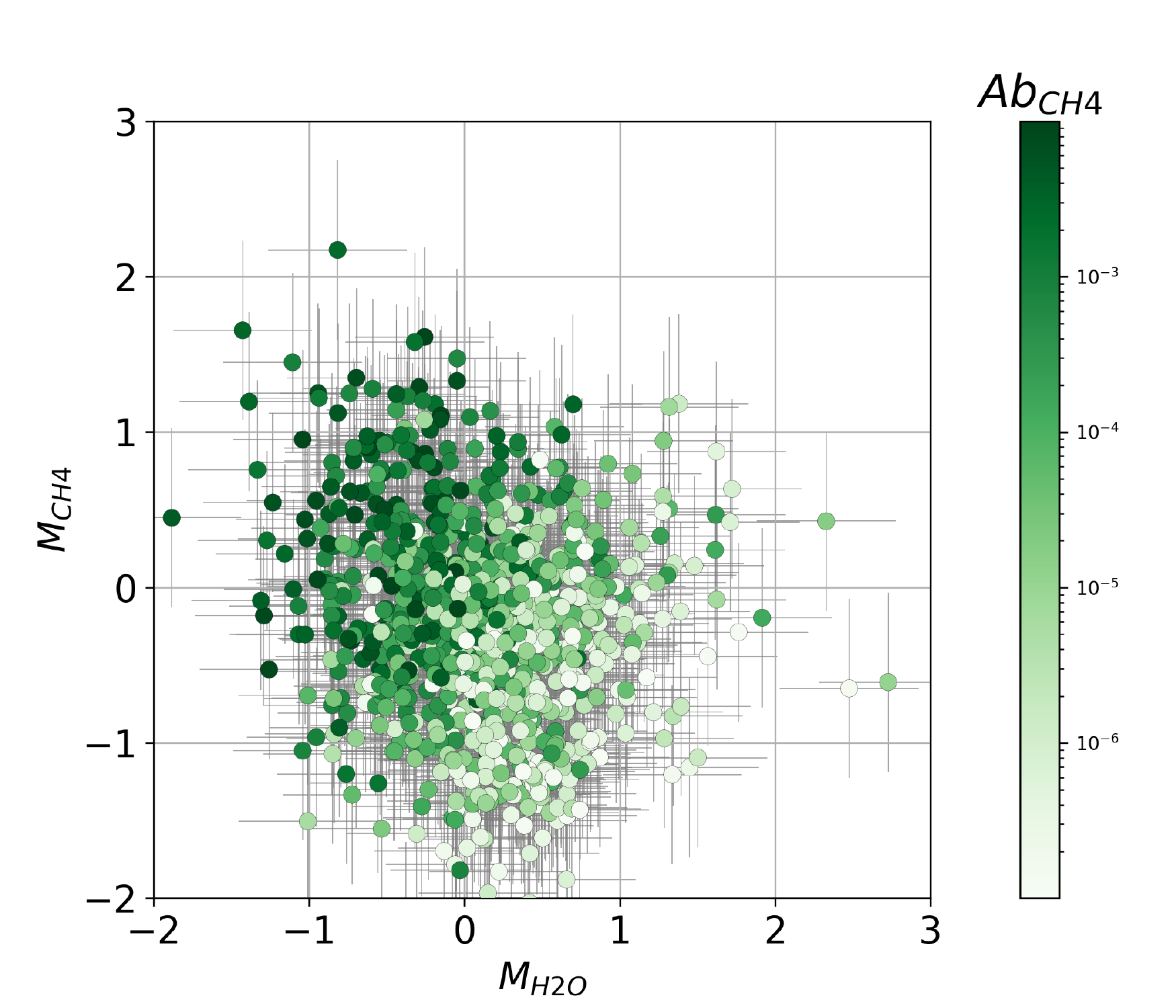}
		\caption{POP-I estimated $M_{H_2O}-M_{CH_4}$ coloured by CH$_4$.}
	\end{subfigure}
	\hfill
	\begin{subfigure}[b]{0.48\textwidth}
		\centering
		\includegraphics[width=\textwidth]{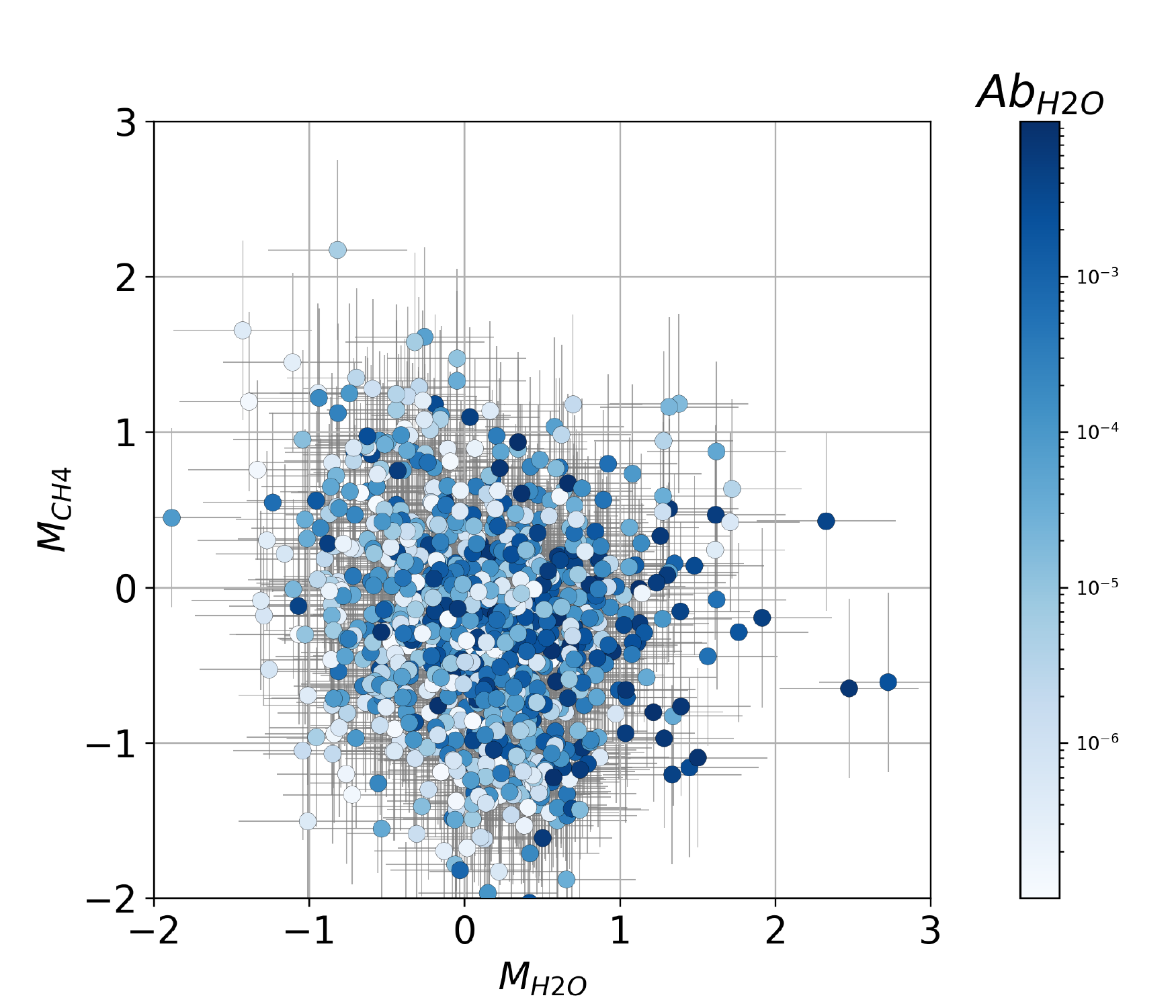}
		\caption{POP-I estimated $M_{H_2O}-M_{CH_4}$ coloured by H$_2$O.}
	\end{subfigure}

	\caption{Diagrams for comparison between $M_{CO_2}$ - $M_{CH_4}$ and  $M_{H_2O}$ - $M_{CH_4}$. In these figures each point represents an observed POP-I planet, and the color scale reflects the input abundances. 
    Grey horizontal and vertical lines are the metric estimated dispersion.  
	By comparing the diagrams on the left with those on the right we can see that planets bearing more CH$_4$ are located on the top left, while the ones bearing more CO$_2$ and H$_2$O are on the bottom right. }
	\label{fig:diagrams}
\end{figure}

The $M_{mol}$ (Sec. \ref{sec:metric}) estimated for the observed POP-I planets are shown in Fig. \ref{fig:diagrams} for different pairs of molecules:  CH$_4$ - CO$_2$ and CH$_4$ - H$_2$O. 
Comparing the top left and right panels in Fig. \ref{fig:diagrams}, we notice from the colour scale that our metric can separate between planets bearing more or less methane (dark and light green coloured dots respectively) or carbon dioxide (dark and light orange coloured dots respectively). The bottom panels, and the bottom-right panel in particular,  show that it is harder to separate planets bearing more or less water (dark and light blue coloured dots respectively). Water data  appear more clustered around the axes' origin than the top row, and the water coloured data points are not as clearly separated according to their colour gradient as the methane or the carbon dioxide data points are. 
A possible explanation is that CH$_4$ and CO$_2$ have strong spectral features, with isolated transmission features in the range $3 \to 4 \; \mu m$ and $4 \to 5 \; \mu m$ respectively, while H$_2$O features are less obvious and frequently overlap with the ones of NH$_3$, that is present in the population \citep{Tinetti2013}.  An alternative explanation is that involving a bias in the metric that affects more strongly the water bands. 

\begin{figure}
	\centering

	\begin{subfigure}{0.38\textwidth}
		\centering
		\includegraphics[width=\textwidth]{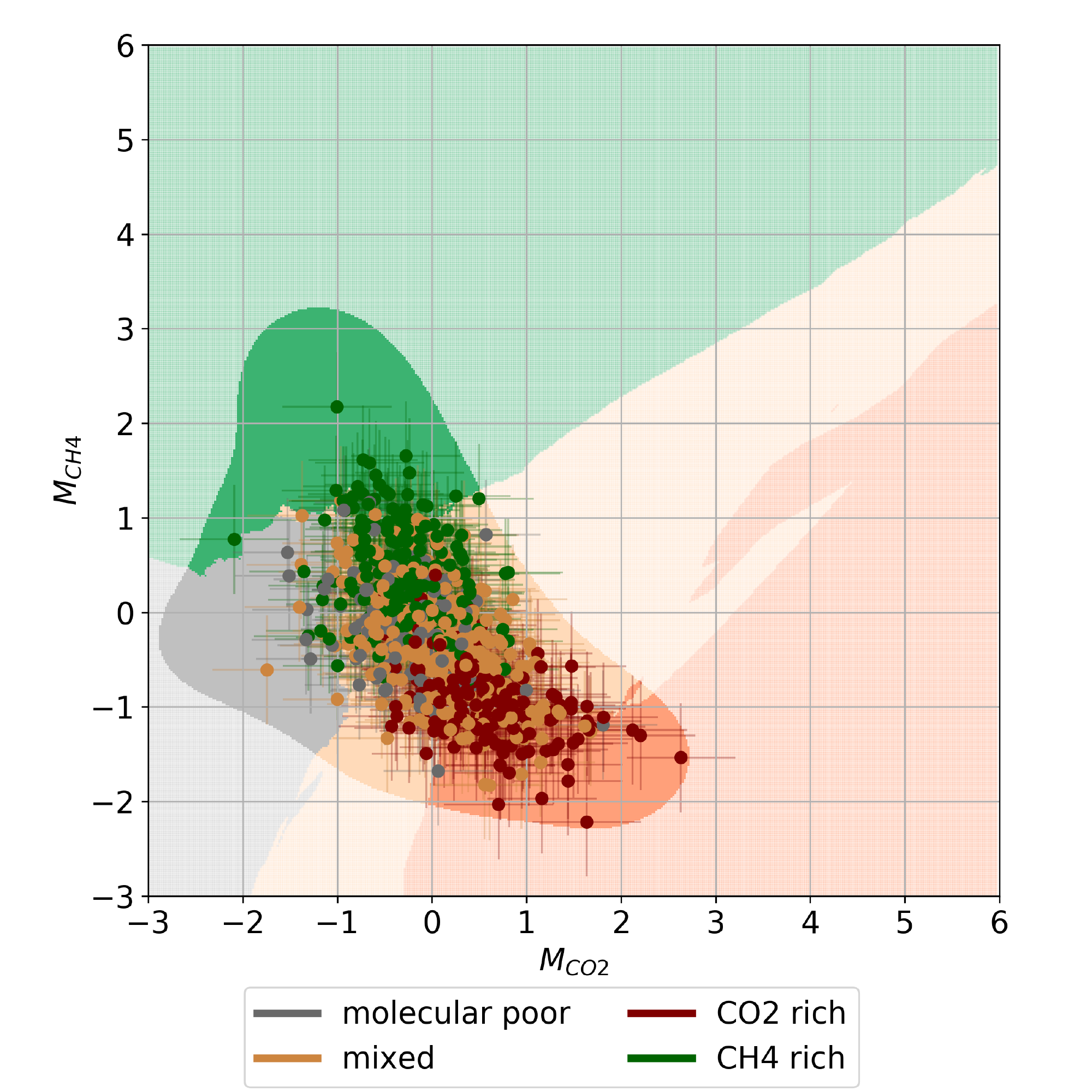}
		\caption{$M_{CO_2}-M_{CH_4}$ - observed spectra.} 
	\end{subfigure}
	\begin{subfigure}{0.38\textwidth}
		\centering
		\includegraphics[width=\textwidth]{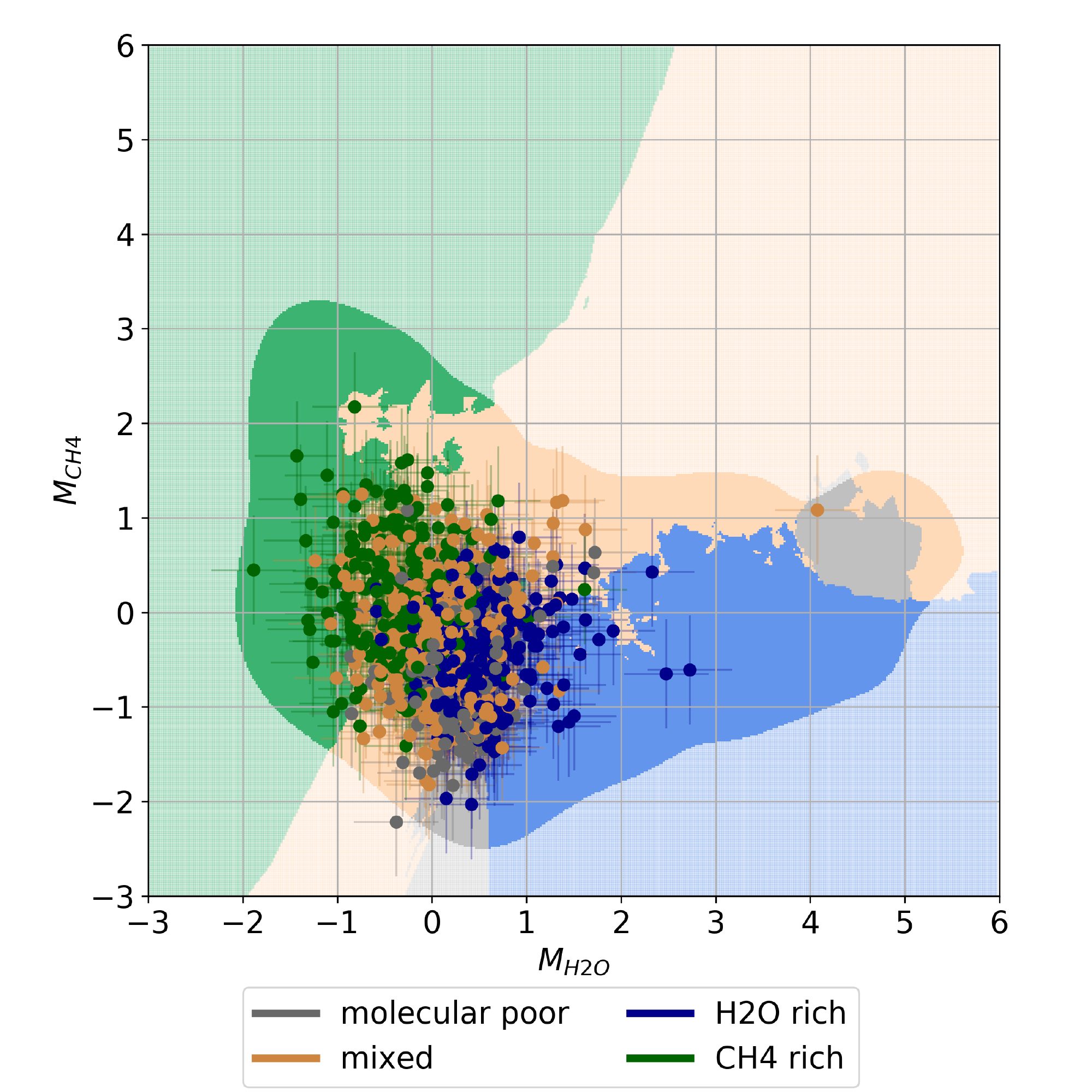}
		\caption{$M_{H_2O}-M_{CH_4}$ - observed spectra.}
	\end{subfigure}

	\caption{KN neighbours analysis results with $k=20$ for CO$_2$-CH$_4$ (left) and H$_2$O-CH$_4$ (right) cases. The superimposed dots are from the POP-I observed spectra and the error bars represent the metric dispersion. Colours correspond to classes described in Tab. \ref{tab:classes}. Grey dots: planets that contain less than $10^{-5}$ in mixing ratio for the considered molecules; green points: planets that contain $10$ times more CH$_4$ than the other molecule and $Ab_{CH_4}>10^{-4}$; red points: planets that hold $10$ times more CO$_2$ than CH$_4$ and $Ab_{CO_2}>10^{-4}$; blue points: planets with $10$ times more H$_2$O than CH$_4$ and $Ab_{H_2O}>10^{-4}$; yellow dots: all the other possible configurations. The same colour scheme applies to the painted region of the diagram, built from the noiseless spectral data. Grey area: planets with low quantities of water and methane; green area: where we expect to have methane rich planets, blue: for water-rich planets; yellow: for mixed atmospheres. The regions best sampled by the noiseless data, as described in Sec. \ref{sec:KNN}, are fully coloured, while other regions are transparent.}
	\label{fig:KNN}
\end{figure}

The diagrams of Fig. \ref{fig:diagrams} are reproduced in Fig. \ref{fig:KNN}, where the data points are now colour coded following the assigned classes ({\it step 1}, Sec. \ref{sec:KNN}) and the background colours, constructed by training the KNN on noiseless spectra ({\it step 2} and {\it 3}, Sec. \ref{sec:KNN}),  serve as reference and calibrated regions in the metric space. It can be noticed that the metric has the desired response from the similarities between the reference regions in Fig. \ref{fig:KNN} with those of Fig. \ref{fig:expected_diagram}, with a clear separation in the metric space. The data points tend to cluster towards the origin of the grid more strongly than the reference regions. This is the effect of the bias, further discussed in Sec. \ref{sec:first_outcomes}.


\begin{figure}
	\centering
	\begin{subfigure}[b]{0.3\textwidth}
		\centering
		\includegraphics[width=\textwidth]{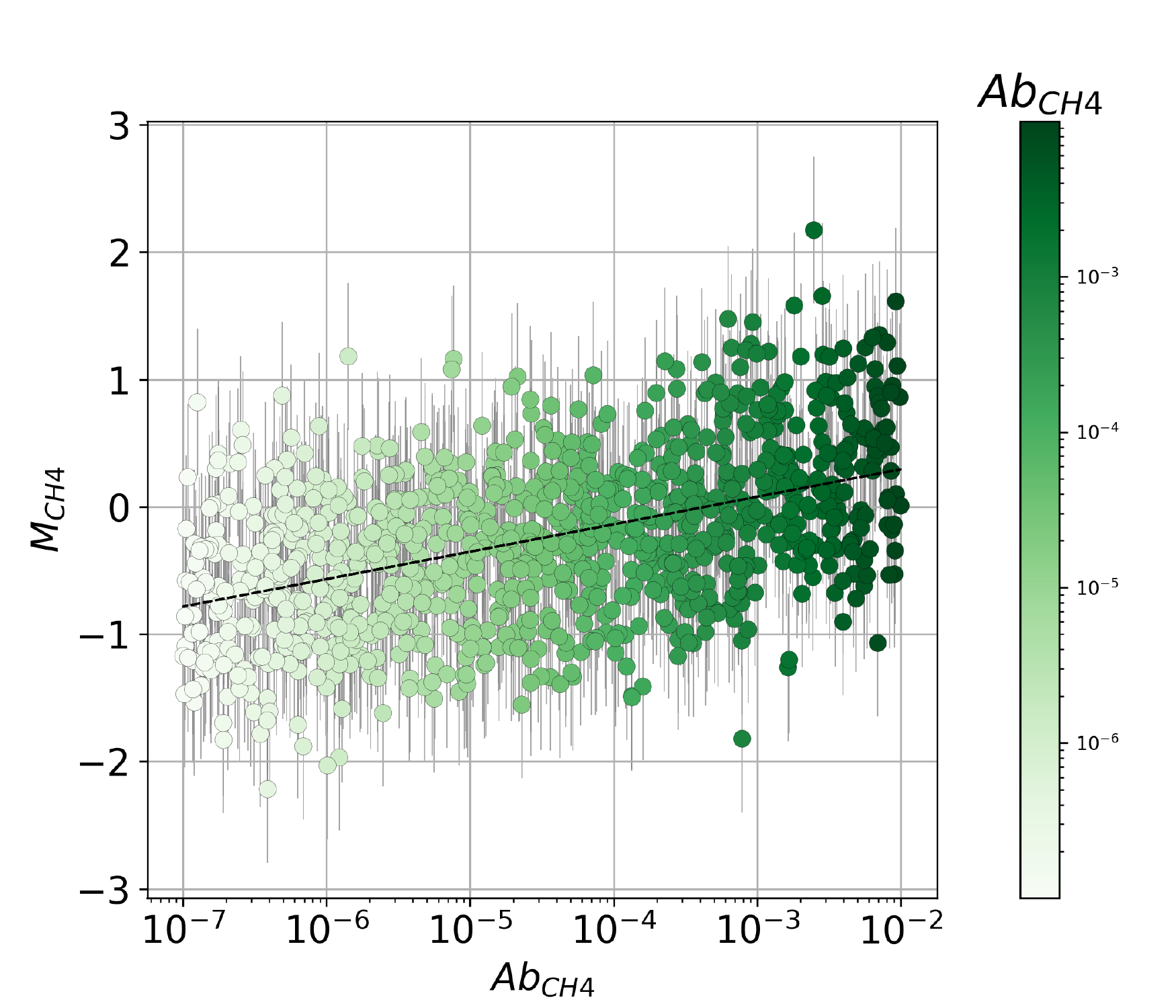}
		\caption{$M_{CH_4}$ versus CH$_4$ true abundance value.}
	\end{subfigure}
	\hfill
	\begin{subfigure}[b]{0.3\textwidth}
		\centering
		\includegraphics[width=\textwidth]{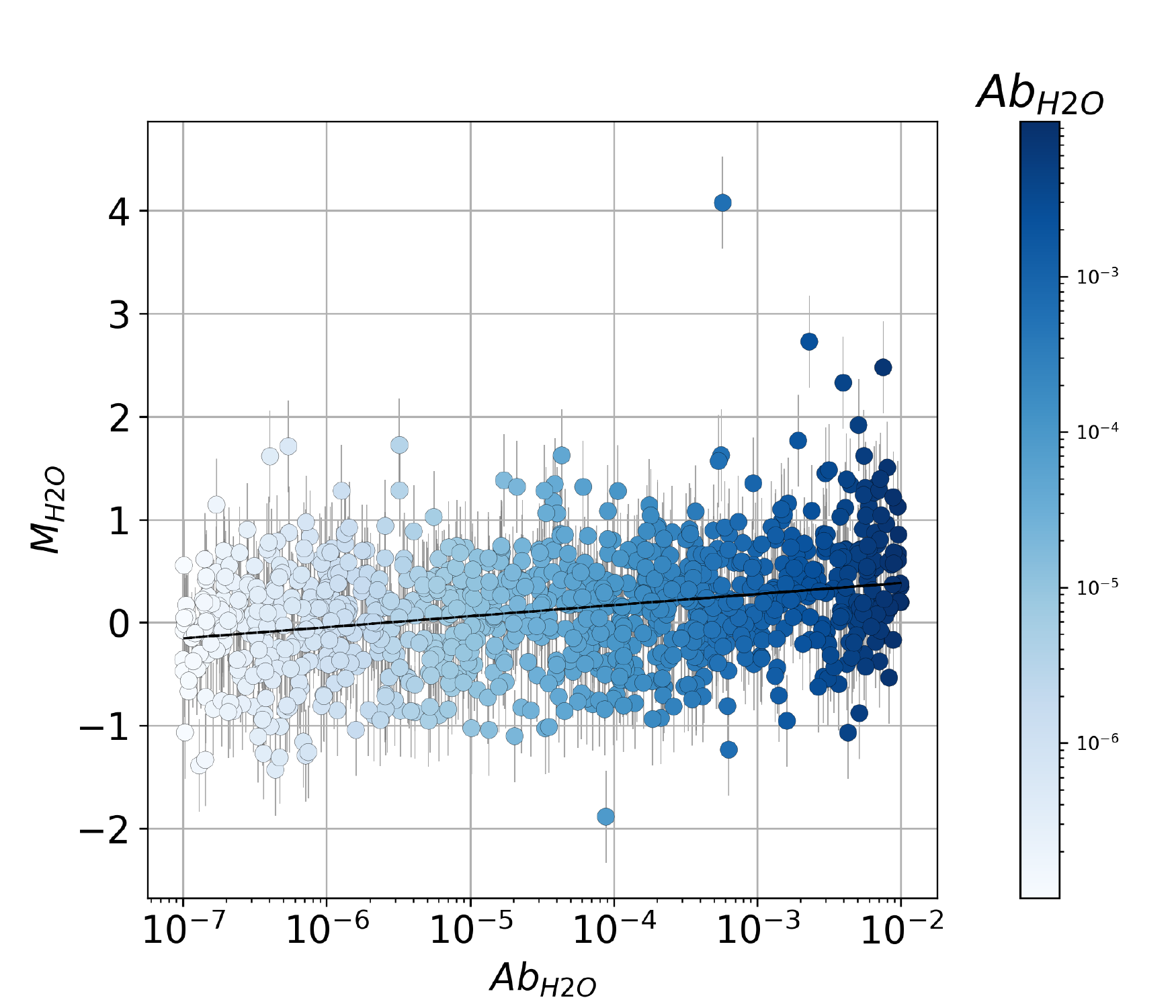}
		\caption{$M_{H_2O}$ versus H$_2$O true abundance value.}
	\end{subfigure}    
	\hfill
	\begin{subfigure}[b]{0.3\textwidth}
		\centering
		\includegraphics[width=\textwidth]{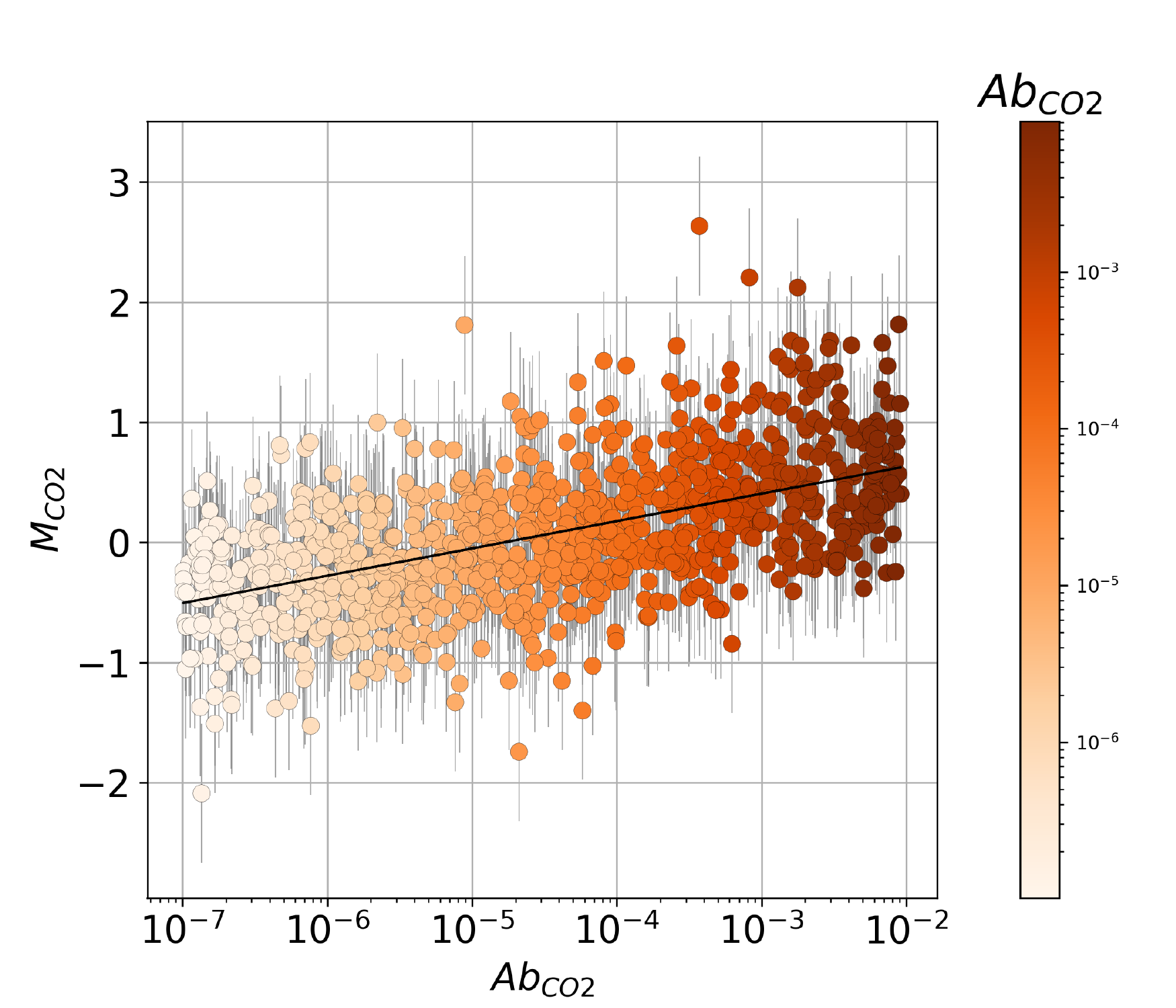}
		\caption{$M_{CO_2}$ versus CO$_2$ true abundance value.}
	\end{subfigure}
	\caption{Comparison between the $M_{mol}$ estimates for each planet and the true molecular abundance value, $Ab_{mol}$,  in the atmospheres. CH$_4$, H$_2$O and CO$_2$ cases are shown in the leftmost, middle and rightmost panel, respectively. Data points with error bars represent POP-I planets. The colour scale gives a visual representation of the molecular abundance in the atmosphere.  A linear fit is shown by the solid black lines in each panel, with coefficients listed in Tab. \ref{tab:cross_diag}. The fitted lines superimposed to the data highlight a positive correlation between the true molecular abundance values and the values estimated by the metric, $M_{mol}$.}
	\label{fig:true_values}
\end{figure}

Fig. \ref{fig:true_values} shows the relation between the metric, $M_{mol}$, estimated on POP-I observed spectra, and the input abundances, $Ab_{mol}$.  The coefficients of the linear trends of $M_{mol}$ vs the logarithm of $Ab_{mol}$ are listed in  Tab. \ref{tab:cross_diag}.
An appreciable trend is detected with log abundances of  CO$_2$ and CH$_4$, while the H$_2$O metric shows only a weak trend with input log abundance. Anti-correlations between e.g., $M_{CH_4}$ - $\log (Ab_{CO_2})$, or $M_{H_2O}$-$\log (Ab_{CH_4})$ are present as we are considering juxtaposed bands to size these molecules, as listed in Tab. \ref{tab:feature_bands}. The logarithmic abundances of H$_2$O and NH$_3$ show similar correlations with $M_{H_2O}$. 
While this is expected, as the two molecules manifest similar spectral shapes, the water sensitivity of the metric to the abundance may also be limited by the noise, by a bias squeezing the metric to small values, or both, and further investigation is required in future work. However, the metric is an estimator for the classification of atmospheres on the basis of their molecular content, and it would be misleading to expect the metric to provide robust estimates of abundances, for which spectral retrieval techniques are more appropriate. These aspects are further discussed in Sec. \ref{sec:relation_abundance} as well as in Sec. \ref{sec:retrieval}, where we show with an example how a retrieval exercise is effective in constraining the input abundances of the molecules considered, water included. 


\begin{table}[]
	\centering
	\caption{Here we report the fitted $C_0$ (top table) and $C_1$ (bottom table) coefficients for $M_{mol} = C_0 \cdot \log(Ab_{mol}) + C_1 $ for all the possible combination of considered molecules. The bands used for $M_{mol}$ are reported in Tab. \ref{tab:feature_bands}.}
	\label{tab:cross_diag}
	\begin{subtable}{\linewidth}\centering
		{
			\caption{$C_0$ coefficients. }
			\begin{tabular}{rrrrr}
				\hline\noalign{\smallskip}
				& {\bf \centering$\log(Ab_{H_2O})$} & {\bf \centering $\log (Ab_{CH_4})$} & {\bf \centering $\log(Ab_{CO_2})$} & {\bf \centering $\log(Ab_{NH_3})$} \\
				\noalign{\smallskip}\hline\noalign{\smallskip}
				$M_{H_2O}$ &  $0.108 (\pm 0.022)$ &   $-0.193 (\pm 0.022)$&  $-0.033 (\pm 0.022)$ & $0.104 (\pm 0.022)$ \\
				$M_{CH_4}$ &  $0.003 (\pm 0.022)$ &  $0.215 (\pm 0.022)$& $-0.258 (\pm 0.022)$ & $0.104 (\pm 0.022)$  \\
				$M_{CO_2}$ &  $-0.094 (\pm 0.022)$&   $-0.057 (\pm 0.022)$& $0.228 (\pm 0.022)$  & $-0.104 (\pm 0.022)$ \\
				\noalign{\smallskip}\hline
			\end{tabular}
			
		}
	\end{subtable}
	
	\break
	\medskip
	
	\begin{subtable}{\linewidth}\centering
		{
			\caption{$C_1$ coefficients. }
			\begin{tabular}{rrrrr}
				\hline\noalign{\smallskip}
				& {\bf \centering$\log(Ab_{H_2O})$} & {\bf \centering $\log (Ab_{CH_4})$} & {\bf \centering $\log(Ab_{CO_2})$} & {\bf \centering $\log(Ab_{NH_3})$} \\
				\noalign{\smallskip}\hline\noalign{\smallskip}
				$M_{H_2O}$ &  $0.599 (\pm 0.102)$ &   $-0.756 (\pm 0.106)$&  $-0.028 (\pm 0.105)$ & $0.590 (\pm 0.105)$   \\
				$M_{CH_4}$ &  $-0.241 (\pm 0.102)$ &  $0.725 (\pm 0.106)$& $-1.419 (\pm 0.105)$ & $0.215 (\pm 0.105)$   \\
				$M_{CO_2}$ &  $-0.357 (\pm 0.102)$&   $-0.202 (\pm 0.106)$& $1.088 (\pm 0.105)$  & $-0.411 (\pm 0.105)$ \\
				\noalign{\smallskip}\hline
			\end{tabular}
			
		}
	\end{subtable}
	
\end{table}

We can use Fig. \ref{fig:true_values} to obtain an estimate of the probability that a molecule $mol$ has abundance in excess of $10^{-4}$, conditioned to the metric being larger than some value $M_{mol,*}$, i.e. $P(Ab_{mol}> 10^{-4} | M_{mol} > M_{mol,*})$.
For this, we can use the well known chain rule for the conditional probability that states that $P(A|B) = P(A \cap B)/P(B)$, where $A$ and $B$ are two separate events. We estimate the number of data points found in a region of the diagrams of Fig. \ref{fig:true_values} where both conditions are satisfied (favourable outcomes) divided by the number of data points for which only the condition $M_{mol} > M_{mol,*}$ is satisfied (total outcomes). 
From POP-I observed spectra, we can obtain a single realisation of $P$. Therefore we simulate 1000 realisations of POP-I observed spectra, using the same input noiseless POP-I population spectra, and randomising the noise realisations. In this way we simulate 1000 realisations of $P$ from which medians and 1-$\sigma$ confidence levels are computed. 


\begin{figure}
	\centering
	\includegraphics[width = 0.9\textwidth]{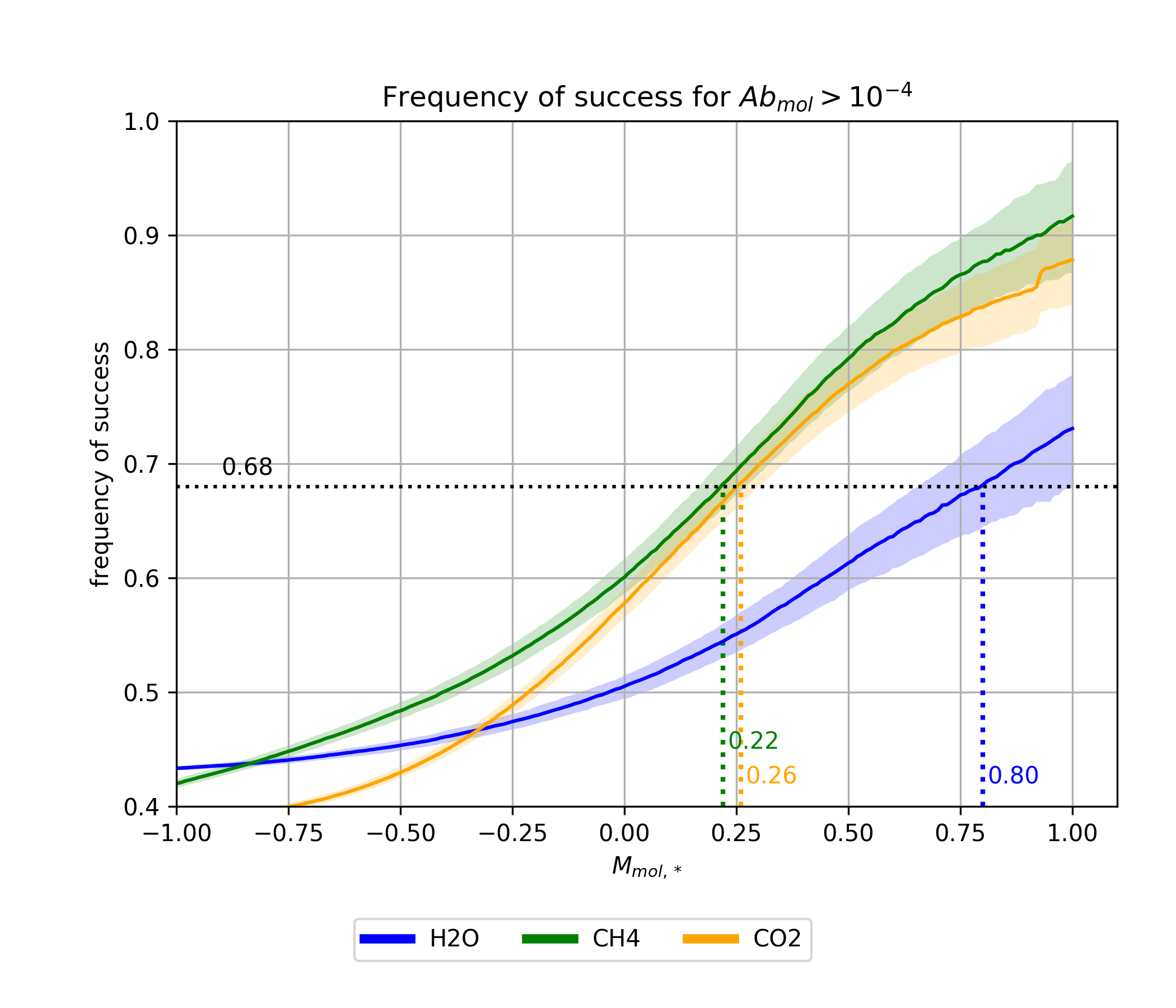}
	\caption{Probability that a molecule $mol$ has abundance in excess of $10^{-4}$, conditioned to the metric being larger than some value $M_{mol,*}$, i.e. $P(Ab_{mol}> 10^{-4} | M_{mol} > M_{mol,*})$. CH$_4$, H$_2$O and CO$_2$ cases are shown by the green, blue and orange lines, respectively. The lines are computed as the median of the probability estimates from 1000 different realisations of the POP-I observed population. 
    The shaded regions are the $1-\sigma$ confidence levels associated with the median probability. Vertical dotted lines mark metric values, $M_{mol,*}$, corresponding to a probability of $68\%$. }
	\label{fig:prob}
\end{figure}

Fig. \ref{fig:prob} suggests that the metric can be used to classify planetary primary atmospheres for the presence of CH$_4$ and CO$_2$, and to a less extent H$_2$O, and atmospheres that are likely missing these molecular contributions. With reference to Fig. \ref{fig:prob}, it can be seen that when $M_{CH_4} \geq 0.5$, the number of planets wrongly classified to have $Ab_{CH_4} > 10^{-4}$ is only $20\%$, or 1 out of 5 are false positives. However, and as expected, the case of water is different, and our metric is not as effective in detecting the presence of water as it is for the other molecules. Even for large values of $M_{H_2O}$, the rate of false positives is close to $40\%$. 


\subsection{Deep and Machine Learning}
\label{sec:deep_results}
The percentages of correct classifications for all considered molecules and for different minimum input abundances are reported in Tab. \ref{tab:Knn-percentages} for KNN, in Tab. \ref{tab:MLP-percentages} for MLP, in Tab. \ref{tab:RFC-percentages} for RFC and in Tab. \ref{tab:SVC-percentages} for SVC. 

\begin{table}[]
	\centering
	\caption{Percentages of correct identifications for the considered molecules and with different thresholds. In each column we report a different minimum $Ab_{mol}$ and in each row a different molecule. The percentages represent how many of the atmospheres have been correctly identified by the algorithm to have at least the specified minimum amount of that molecule, and therefore they represent the algorithm accuracy. Each ML algorithm has been trained on POP-III and tested on POP-I.}
	\label{tab:deeplearning}
	\begin{subtable}{\textwidth}
		\centering
		\caption{KNN percentages of success to identify spectra bearing different minimum amounts of molecules.}
		\label{tab:Knn-percentages}
		\begin{tabular}{cccc}
			\hline\noalign{\smallskip}
			{\bf \centering Molecule} & {\bf \centering $Ab_{mol}>10^{-5}$ [$\%$]} & {\bf \centering $Ab_{mol}>10^{-4}$ [$\%$]} & {\bf \centering $Ab_{mol}>10^{-3}$ [$\%$]} \\
			\noalign{\smallskip}\hline\noalign{\smallskip}         
			CH$_4$ & 79 & 83 & 85\\
			CO$_2$ & 77 & 79 & 82\\
			H$_2$O & 64 & 71 & 82 \\
			NH$_3$ & 75 & 82 & 84 \\
			\noalign{\smallskip}\hline
			
		\end{tabular}
	\end{subtable}
	
	\begin{subtable}{\textwidth}
		\centering
		\caption{MLP percentages of success to identify spectra bearing different minimum amounts of molecules.}
		\label{tab:MLP-percentages}
		\begin{tabular}{cccc}
			\hline\noalign{\smallskip}
			{\bf \centering Molecule} & {\bf \centering $Ab_{mol}>10^{-5}$ [$\%$]} & {\bf \centering $Ab_{mol}>10^{-4}$ [$\%$]} & {\bf \centering $Ab_{mol}>10^{-3}$ [$\%$]} \\
			\noalign{\smallskip}\hline\noalign{\smallskip}         
			CH$_4$ & 78 & 85 & 87\\
			CO$_2$ & 77 & 81 & 83\\
			H$_2$O & 70 & 76 & 84 \\
			NH$_3$ & 80 & 86 & 87 \\
			\noalign{\smallskip}\hline
		\end{tabular}
	\end{subtable}
	
	\begin{subtable}{\textwidth}
		\centering
		\caption{RFC percentages of success to identify spectra bearing different minimum amounts of molecules.}
		\label{tab:RFC-percentages}
		\begin{tabular}{cccc}
			\hline\noalign{\smallskip}
			{\bf \centering Molecule} & {\bf \centering $Ab_{mol}>10^{-5}$ [$\%$]} & {\bf \centering $Ab_{mol}>10^{-4}$ [$\%$]} & {\bf \centering $Ab_{mol}>10^{-3}$ [$\%$]} \\
			\noalign{\smallskip}\hline\noalign{\smallskip}         
			CH$_4$ & 77 & 82 & 87\\
			CO$_2$ & 76 & 79 & 83\\
			H$_2$O & 69 & 74 & 82 \\
			NH$_3$ & 78 & 85 & 87 \\
			\noalign{\smallskip}\hline
		\end{tabular}
		
	\end{subtable}
	
	\begin{subtable}{\textwidth}
		\centering
		\caption{SVC percentages of success to identify spectra bearing different minimum amounts of molecules.}
		\label{tab:SVC-percentages}
		\begin{tabular}{cccc}
			\hline\noalign{\smallskip}
			{\bf \centering Molecule} & {\bf \centering $Ab_{mol}>10^{-5}$ [$\%$]} & {\bf \centering $Ab_{mol}>10^{-4}$ [$\%$]} & {\bf \centering $Ab_{mol}>10^{-3}$ [$\%$]} \\
			\noalign{\smallskip}\hline\noalign{\smallskip}         
			CH$_4$ & 79 & 86 & 89\\
			CO$_2$ & 79 & 83 & 84\\
			H$_2$O & 69 & 78 & 84 \\
			NH$_3$ & 81 & 87 & 87 \\
			\noalign{\smallskip}\hline
		\end{tabular}
		
	\end{subtable}
	
\end{table}

Tab. \ref{tab:deeplearning} shows that for all Deep and Machine Learning algorithms, the percentages of success in identifying the presence of molecules inside the atmosphere grow with the minimum molecular abundances that we set as a threshold for the classification. 
While this is expected, it may come as a surprise that in general these algorithms appear to be effective in detecting the presence of all individual molecules with a relatively small fraction of false positives (about $30\%$ or smaller) even at low abundances. 
This is perhaps because ML algorithms learn to classify atmospheres by recognising spectral shapes.
These algorithms performances can be to a certain level independent of the molecules considered, as long as the training set contains sufficiently diverse spectra to allow a secure identification, including water in the presence of ammonia or biases,  that is where our metric shows its more severe weaknesses.
We also notice from Tab. \ref{tab:deeplearning} that KNN, MLP, RFC and SVC show comparable overall performance, and that CH$_4$ and CO$_2$ are the most straightforward molecules to identify in Tier 1 planetary spectra. 

A comparison between these results and our metric is presented in Sec. \ref{sec:deep_discussion}.

\section{Discussion}
\label{sec:discuss}
In this section, we discuss the metric results shown in Sec. \ref{sec:first_results}. We first discuss the bias  (Sec. \ref{sec:first_outcomes}), then we focus on the metric characteristics, such as the relation between the metric estimates and the input molecular abundances (Sec. \ref{sec:relation_abundance}) and the detection limits (Sec. \ref{sec:detection_limit}). Then we compare the metric performance with a spectral retrieval (Sec. \ref{sec:retrieval}), and with Deep and Machine Learning algorithms (Sec. \ref{sec:deep_discussion}). 

\subsection{Metric bias} \label{sec:first_outcomes}

\begin{figure}
	\centering

	\centering
	\begin{subfigure}{0.38\textwidth}
		\centering
		\includegraphics[width=\textwidth]{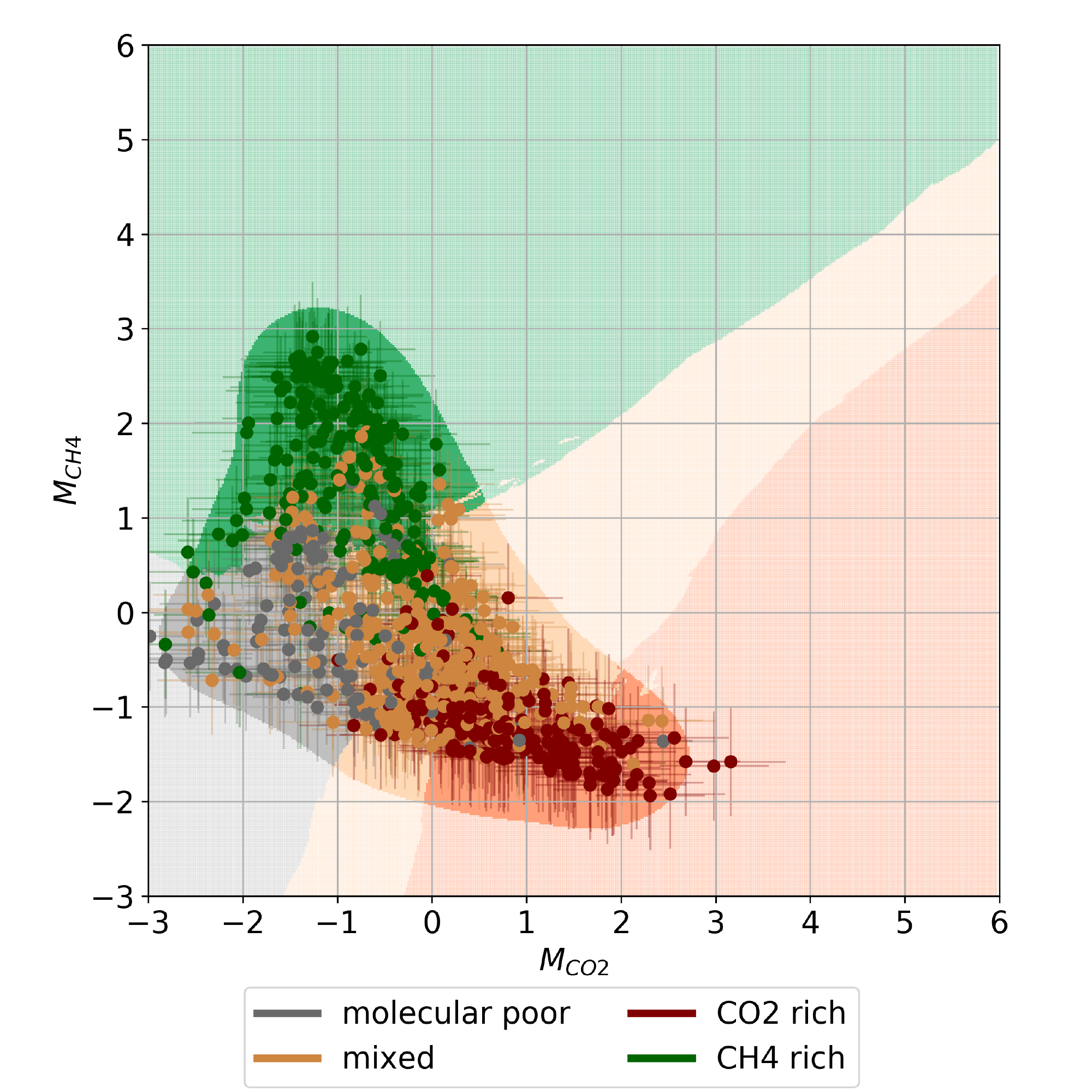}
		\caption{$M_{CO_2}-M_{CH_4}$ - noiseless spectra.} 
	\end{subfigure}
	\begin{subfigure}{0.38\textwidth}
		\centering
		\includegraphics[width=\textwidth]{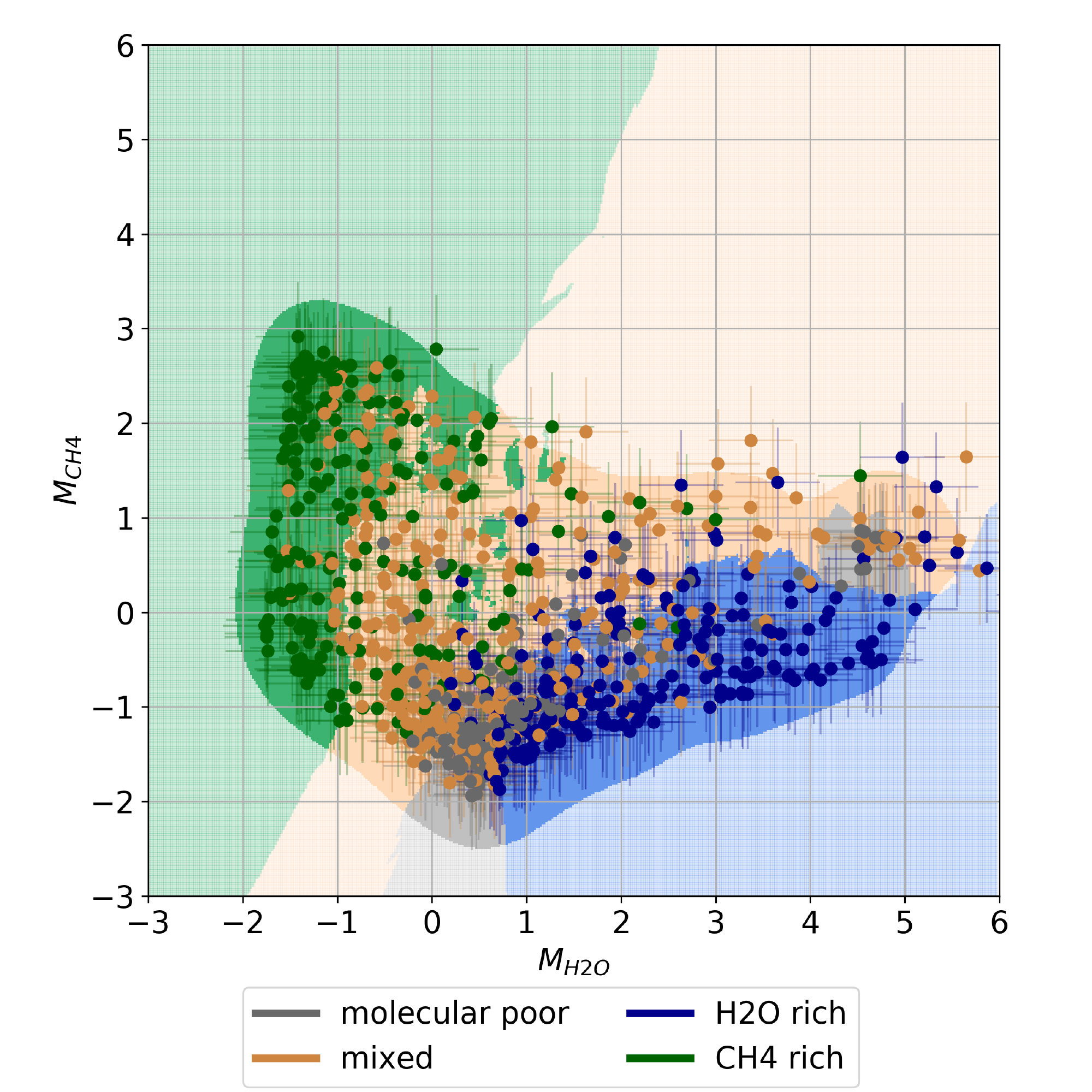}
		\caption{$M_{H_2O}-M_{CH_4}$ - noiseless spectra.}
	\end{subfigure}

	\caption{This figure is the equivalent of Fig. \ref{fig:KNN}, but the superimposed dots are now from the POP-I noiseless spectra, and the errorbars represent the metric dispersion on the spectra before the application of \ARIEL's observing noise. The parameter space area best sampled by the noiseless data is now well filled with the dots.}
	\label{fig:KNN_clean}
\end{figure}

The KNN analysis discussed earlier and shown in Fig. \ref{fig:KNN} is trained on POP-I noiseless spectra, and the data-points shown in that figure are obtained estimating the metric on POP-I observed spectra, as described in Sec. \ref{sec:KNN}. To verify if the metric is biased, the KNN analysis is repeated with data-points obtained estimating the metric on POP-I noiseless spectra. This is shown in  Fig. \ref{fig:KNN_clean} that should be compared with Fig. \ref{fig:KNN}. The background colours are very similar in either cases, with small variations due to the training process that selects randomly 70\% POP-I noiseless examples. In absence of biases, we expect the distribution of observed data-points to be that of noise-less data-points, convolved with the distribution of the noise. However, it can be noticed from the comparison of the two figures, that the distribution of the observations is more clustered towards the origin of the coordinate axes, compared to noiseless data-points. This is a consequence of the bias introduced by the metric normalisation discussed in Sec. \ref{sec:metric}: normalisation is required such that the metric response is insensitive to the atmospheric scale height, and sensitive only to the presence of molecular signatures, at the cost of biasing the estimator. We should additionally point out that  Fig. \ref{fig:prob} results are also affected by the bias. The observing noise reduces the $M_{mol}$ average estimates, and therefore for smaller observing noise, the three coloured lines in the figure are shifted to the right, and the $68 \%$ of success corresponds to higher $M_{mol}$ values.

The work presented here demonstrates that the metric we have designed is a powerful tool capable of revealing the presence of a molecule in an atmosphere and that the prediction is independent of the type of the planet and its basic parameters (such as temperature, radius,  and pressure) within the limits explored here. However, this comes at the cost of biasing the estimator by a quantity that depends on the instrumental noise as discussed in Sec. \ref{sec:metric}. Provided that the metric can be de-biased, it can be used in a predictive way where an observation (along with its dispersion estimate) can be compared to the calibrated (trained) metric space to infer the possible molecular content of the target.  Because instrumental noise can be well characterised, it would be possible to de-bias the metric estimator. This requires a detailed noise analysis, taking into account the uncertainties on the noise estimates, which is beyond the scope of this paper. In the rest of this section we focus on what we can learn from this kind of analysis provided that the metric can be de-biased, and we leave to future work a detailed study on how this de-biasing can be secured.

\subsection{Relation with the input abundances}\label{sec:relation_abundance}

We see in Fig. \ref{fig:true_values} that the correlation between $M_{mol}$ and $\log(Ab_{mol})$ is in general not strong enough to quantify the input molecular abundances. This is because atmospheric spectra are made of complex non-linear contributions from all the molecules. Therefore, a method based only on spectral shapes (i.e., this metric), is inadequate to quantify molecular abundances. 
However, the goal of this metric, provided that the bias can be removed, is not to assess the abundance of a certain species in the planet atmosphere, but only its possible presence, avoiding the use of spectral retrieval techniques, that may not be indicated for Tier 1 data.

Focusing on Tab. \ref{tab:cross_diag} and looking at the coefficients fitted for $M_{H_2O}$ over $\log(Ab_{H_2O})$ and over $\log(Ab_{NH_3})$ we may infer that the metric may not be effective to distinguish between water and ammonia. However, the degeneracy can be broken by performing a spectral retrieval if the target was observed at \ARIEL\ Tier 2 SNR, as shown in an example in sec \ref{sec:retrieval}.
This population analysis is based on the study of spectral shapes only, and it does not make use of parameters such as planetary mass, radius and temperature. Although it has proven difficult to distinguish between water and ammonia with this metric, using some knowledge of planetary properties may help us to disentangle the two molecules in a future work; for example, while a Neptune can hold ammonia, a Hot Jupiter planet is not expected to.  
One of the goals of  Tier 1 is to identify targets with interesting spectra to be re-observed in higher SNR Tiers. From this point of view, even if the metric cannot clearly separate between water and ammonia, it can suggest the presence of interesting molecules in the spectrum.  This can in turn be used to make informed decisions about targets to be selected for further studies.


\subsection{Metric detection limit}\label{sec:detection_limit}

To explore the detection limit of molecules by the metric, we examine the molecular poor/spectral flat region of Fig. \ref{fig:expected_diagram}. A planet spectrum would be found in that region because of i) clouds, ii) a low temperature (i.e. small scale height), iii) low molecular abundances or a combination of the three.  In all cases, the spectrum is expected to be featureless, i.e. flat. Point iii) is defined from input abundances smaller than $10^{-5}$ (Tab. \ref{tab:classes}). The metric detection limit can then be investigated by removing flat spectra before training the KNN, by rising before training the molecular poor spectra threshold to above $10^{-5}$, and by monitoring the KNN classification results. As the threshold increases, we expect the KNN to begin failing the molecular poor/flat classification when spectra can no longer be considered flat.

We perform the KNN training on the noiseless spectra of both POP-I and POP-II, the latter containing only CH$_4$ and H$_2$O, the former containing all molecules considered in this work. 
Each noiseless spectrum has its associated observed spectrum. Flat spectra are identified on observed spectra, and the corresponding noiseless spectra are ignored in the KNN training. 

\begin{figure}
	\centering
	\includegraphics[width = \textwidth]{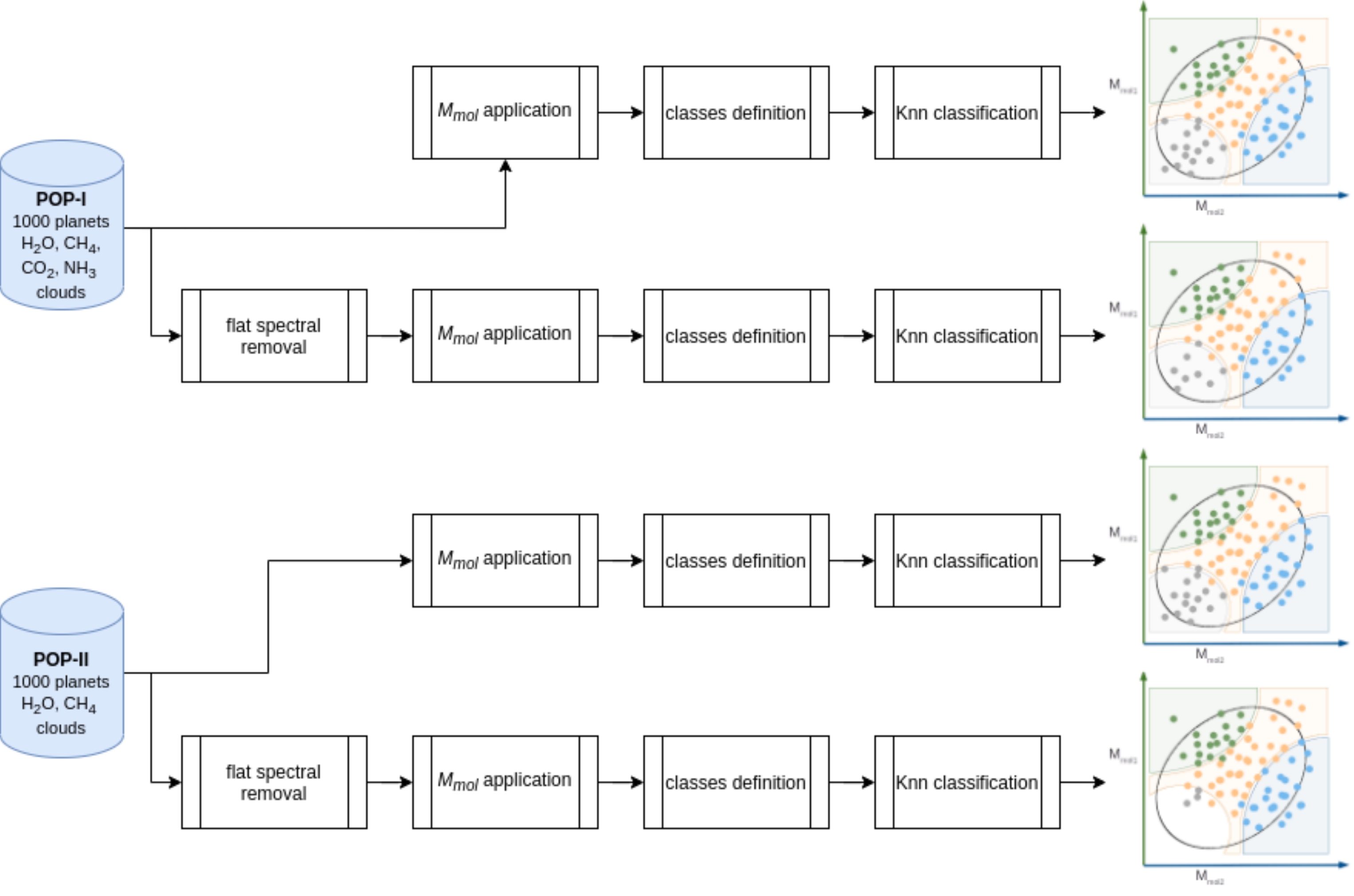}
	\caption{The figure shows the strategy adopted to identify the molecular detection limit for the developed metric. Starting from \textit{POP-I}, we classify the planets as described in Sec. \ref{sec:KNN}. Without removing the flat spectra from the population, we would end up with the same results described in Fig. \ref{fig:knn_summary}; by contrast, if we remove flat spectra, we end up with similar results but with fewer molecular poor planets, because even without flat spectra atmospheres, there will be planets bearing molecules different from the couple investigated by the plot. Different is the case of \textit{POP-II}: here we have only two molecules in the population, and therefore if we remove the flat spectra planets, we will end with no molecular poor atmospheres.}
	\label{fig:flat_summary}
\end{figure}

The motivation behind using POP-II is as follows. If we have a population containing only CH$_4$ and H$_2$O and we properly remove all planets with a flat spectrum, there should be no targets left with non-detectable molecular features. 
In the case of POP-I, however, we do not expect all the planets with $Ab_{CH_4}$ and $Ab_{H_2O}< 10^{-5}$  to be flat, because other molecules (CO$_2$ and NH$_3$) can show features. Therefore, the flat spectra removal procedure will not empty the molecular poor planets class in this population. Using POP-II instead, we expect that, after removing all flat planets, there will not be molecular poor atmospheres anymore. 
The procedure is summarised in Fig. \ref{fig:flat_summary}.

\begin{figure}
	\centering
	\begin{subfigure}[b]{0.31\textwidth}
		\centering
		\includegraphics[width=\textwidth]{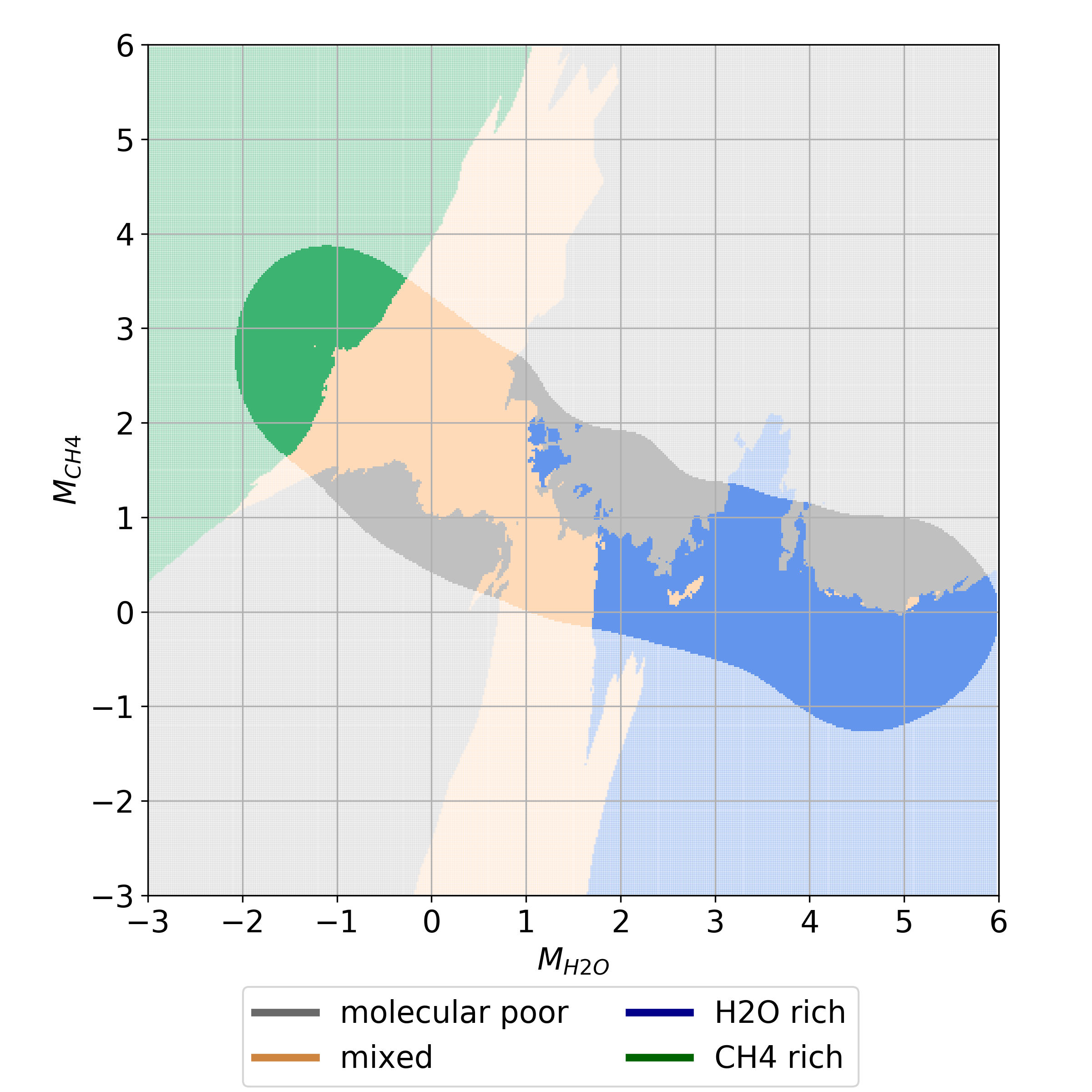}
		\caption{KNN classification map for CH$_4$ and H$_2$O, including flat spectra. Molecular poor planets defined as $Ab_{CH_4}$ and $Ab_{H_2O} < 10^{-5}$. \label{fig:KNN_masked_a}}
	\end{subfigure}
	\hfill
	\begin{subfigure}[b]{0.31\textwidth}
		\centering
		\includegraphics[width=\textwidth]{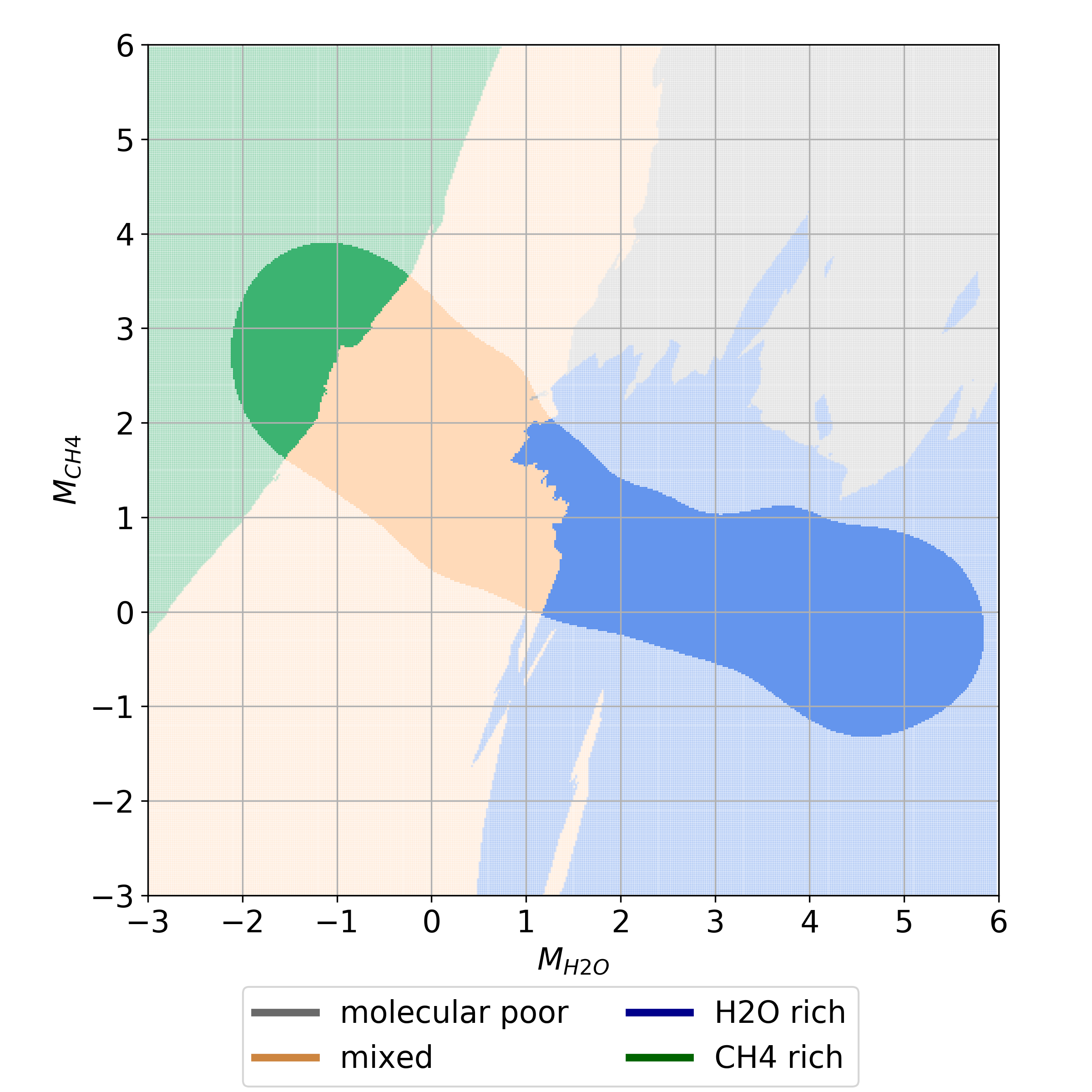}
		\caption{KNN classification map for CH$_4$ and H$_2$O without flat spectra. Molecular poor planets defined as $Ab_{CH_4}$ and $Ab_{H_2O} < 10^{-5}$. \label{fig:KNN_masked_b}}
	\end{subfigure}
	\hfill
	\begin{subfigure}[b]{0.31\textwidth}
		\centering
		\includegraphics[width=\textwidth]{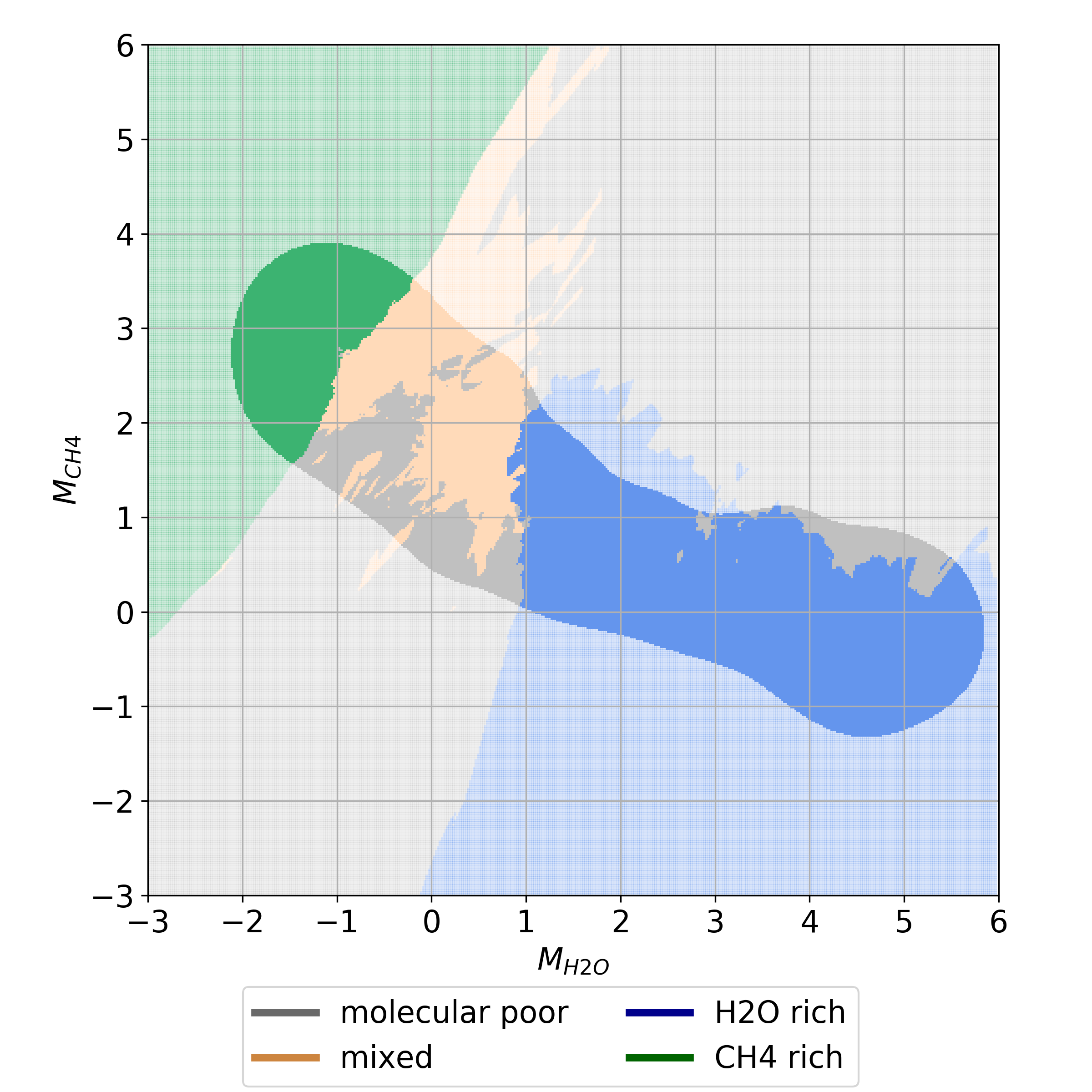}
		\caption{KNN classification map for CH$_4$ and H$_2$O without flat spectra. Molecular poor planets defined as $Ab_{CH_4}$ and $Ab_{H_2O} < 10^{-4}$. \label{fig:KNN_masked_c}}
	\end{subfigure}
	
	\caption{ KNN analysis  for the  POP-II population, considering the full data set (left) and excluding  flat spectra (centre and right). The diagrams are obtained following the bottom branches of Fig. \ref{fig:knn_summary}: we used the noiseless planetary spectra to classify the metric space and to select the best sampled regions.}
	\label{fig:KNN_masked}
\end{figure}

\begin{figure}
	\centering
		\begin{subfigure}[b]{0.48\textwidth}
		\centering
		\includegraphics[width=\textwidth]{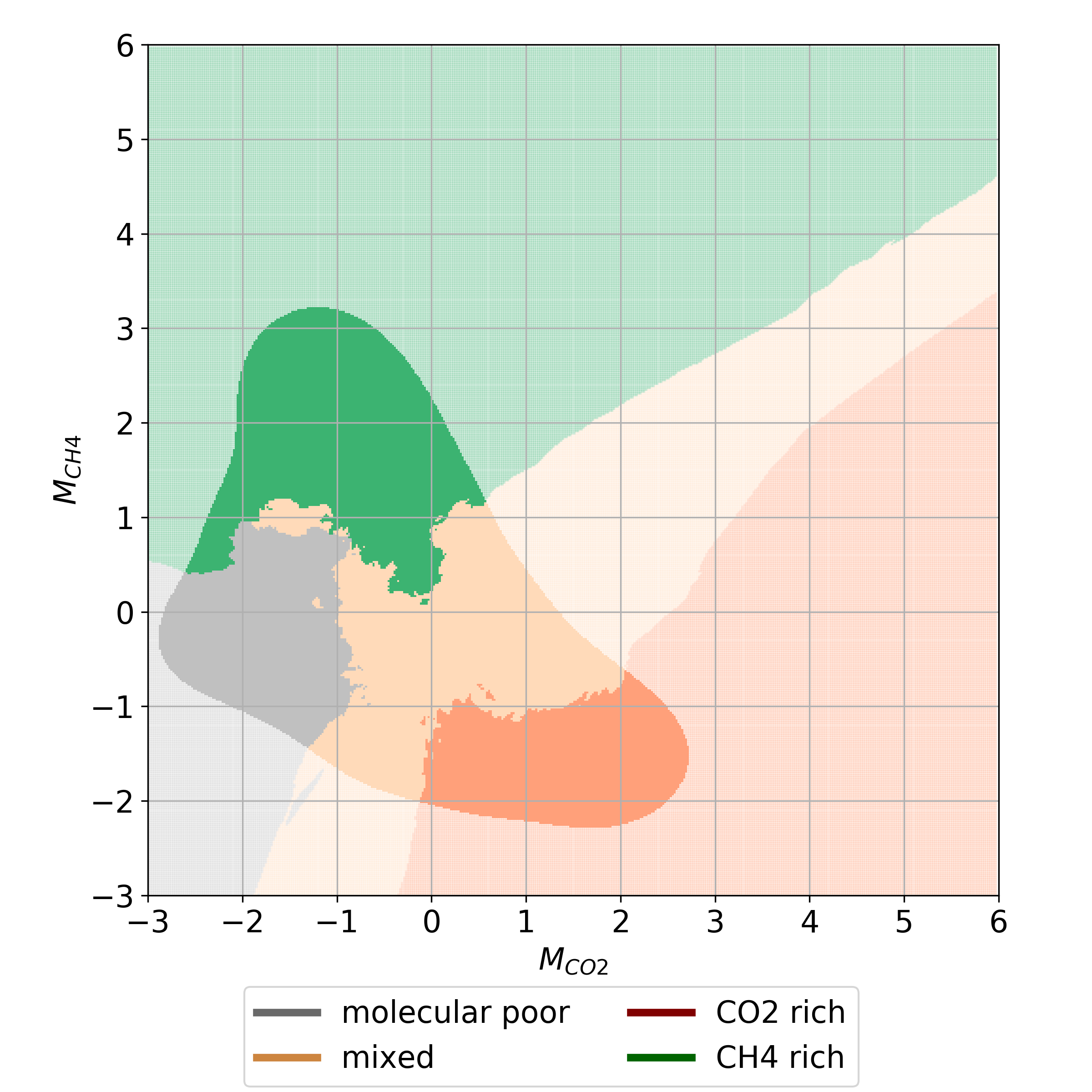}
		\caption{KN neighbours for POP-I population without flat spectra. CH$_4$-CO$_2$ case. 
        }
	\end{subfigure}
	\hfill	
	\begin{subfigure}[b]{0.48\textwidth}
		\centering
		\includegraphics[width=\textwidth]{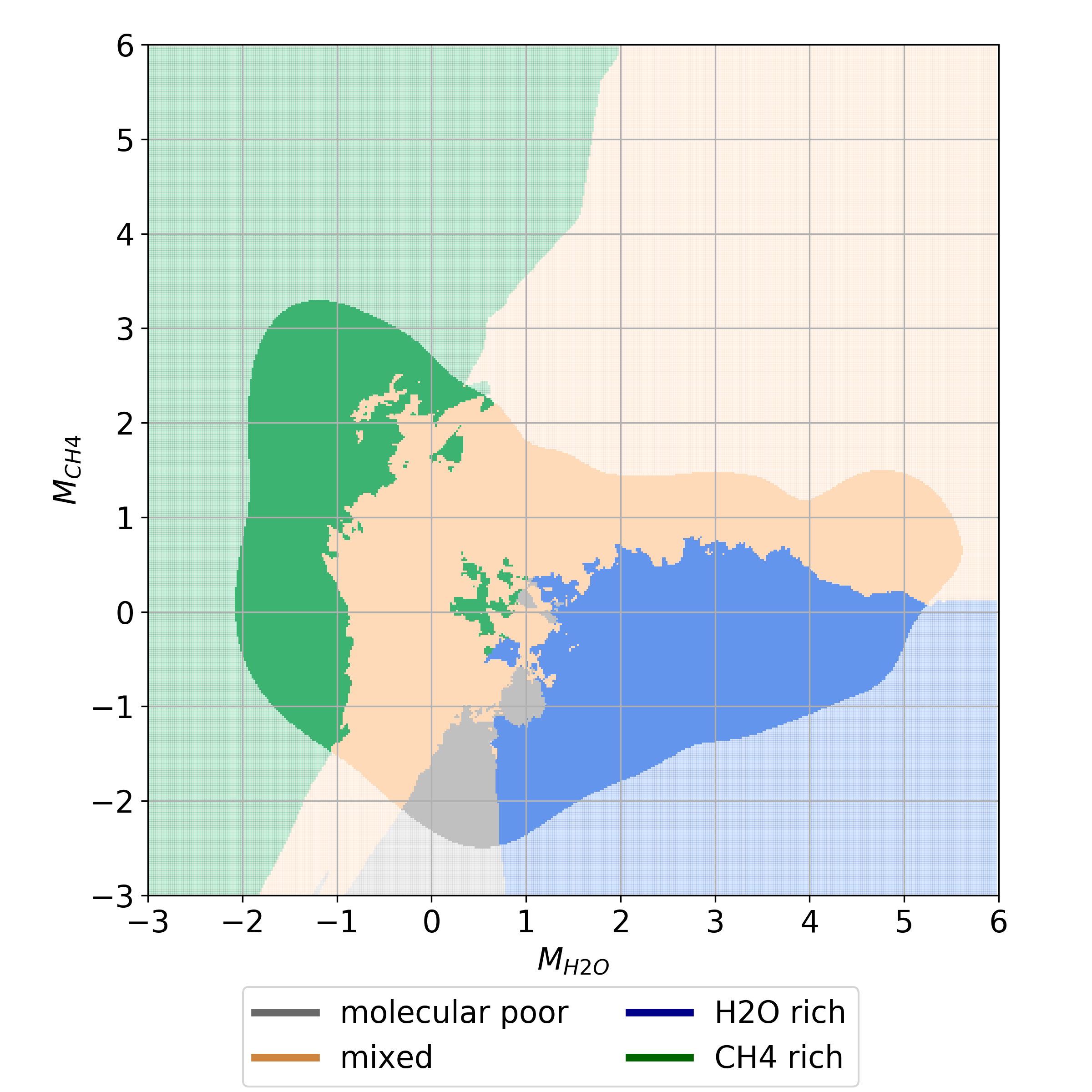}
		\caption{KN neighbours for POP-I population without flat spectra. CH$_4$-H$_2$O case. 
        }
	\end{subfigure}
	
	\caption{The figure shows the population POP-I where the flat spectra have been masked. On the left is reported the case of CH$_4$ and CO$_2$ and on the right the case of CH$_4$ and H$_2$O. The diagrams are obtained following the bottom branches of Fig. \ref{fig:knn_summary}: we used the noiseless planetary spectra to classify the parameter space and to select the best sampled areas. 
    Comparing this figure with  Fig. \ref{fig:KNN}, we notice that the ``molecular poor'' area is still present because even if there are no CO$_2$ and CH$_4$ in the planet atmosphere, there could be NH$_3$ and H$_2$O having features (left case) or if there are no H$_2$O and CO$_2$ there could be NH$_3$ and CO$_2$ (right case).}
	\label{fig:KNN_masked2}
\end{figure}

The outcome of this analysis is shown for   POP-I and POP-II in respectively Fig. \ref{fig:KNN_masked} and  Fig. \ref{fig:KNN_masked2}. Only the calibrated regions are shown and data-points have been omitted for clarity. 
Fig. \ref{fig:KNN_masked_a} shows the POP-II KNN analysis with all planets and planetary classes of Tab. \ref{tab:classes}, in Fig. \ref{fig:KNN_masked_b} the KNN is trained removing flat spectra from the training set, and  in Fig. \ref{fig:KNN_masked_c} the training is done removing flat spectra first, and rising the threshold of molecular poor spectra from $Ab_{mol}<10^{-5}$ to $Ab_{mol}<10^{-4}$.
We notice that Fig. \ref{fig:KNN_masked_b} shows no molecular poor atmosphere after excluding spectrally flat cases. This confirms that our metric is able to separate the more complex atmospheres from the flat ones in the simple case of only two molecules. By contrast, Fig. \ref{fig:KNN_masked_c} still shows a grey area, signifying that atmospheres with $10^{-5} < Ab_{mol} < 10^{-4}$ cannot be considered flat. This can be interpreted as a molecular detection limit. We also notice from the figure that these spectra populate the bottom left corner of the best sampled area of the diagram, meaning that they are classified as having the smallest spectral features of the samples. This confirms the relation between the metric and the molecule abundance. The detection limit is expected to improve in Tier 2 observations, and  \citet{Changeat2020} find that the detection limit using spectral retrieval techniques on Tier 2 is about two orders of magnitude smaller compared to that of the metric.

In Fig. \ref{fig:KNN_masked2} we remove all flat spectra from the planetary population POP-I and we report the results of KNN analysis.  Here we see that, as expected, while removing all flat spectra from POP-II does also remove all molecular-poor instances, the same does not occur in POP-I. In this case, molecular-poor spectra in any two molecules,  such as CH$_4$-CO$_2$ or CH$_4$-H$_2$O, may appear non-flat because of the presence of the other two molecules, i.e. NH$_3$-H$_2$O or NH$_3$-CO$_2$, respectively. 

\subsection{ Input abundances retrieval}
\label{sec:retrieval}

We compare here two atmospheric retrievals of the same planet observed both in Tier 1 and in Tier 2. This exercise has two goals: 
\begin{enumerate}
    \item to confirm that a spectral retrieval is capable of disentangling water and ammonia, and to constrain the atmospheric composition of POP-I targets observed in Tier 2 with \ARIEL;
    \item to show that even though it is possible to perform a spectral retrieval on Tier 1 data for some selected planets, its performance is comparable with that of the metric.  
\end{enumerate}

From the POP-I planets, we select one that has water and ammonia in high abundances, low cloud presence, high temperature and a diameter larger than Jupiter's. Such selection will help us to investigate the capability of Tier 2 observed data (simulated as described in Sec. \ref{sec:planetary population}) to break the water-ammonia degeneracy, as well as to estimate the uncertainties from a retrieval using Tier 1 observed data only.

\begin{figure}
	\centering
	\includegraphics[width=\textwidth]{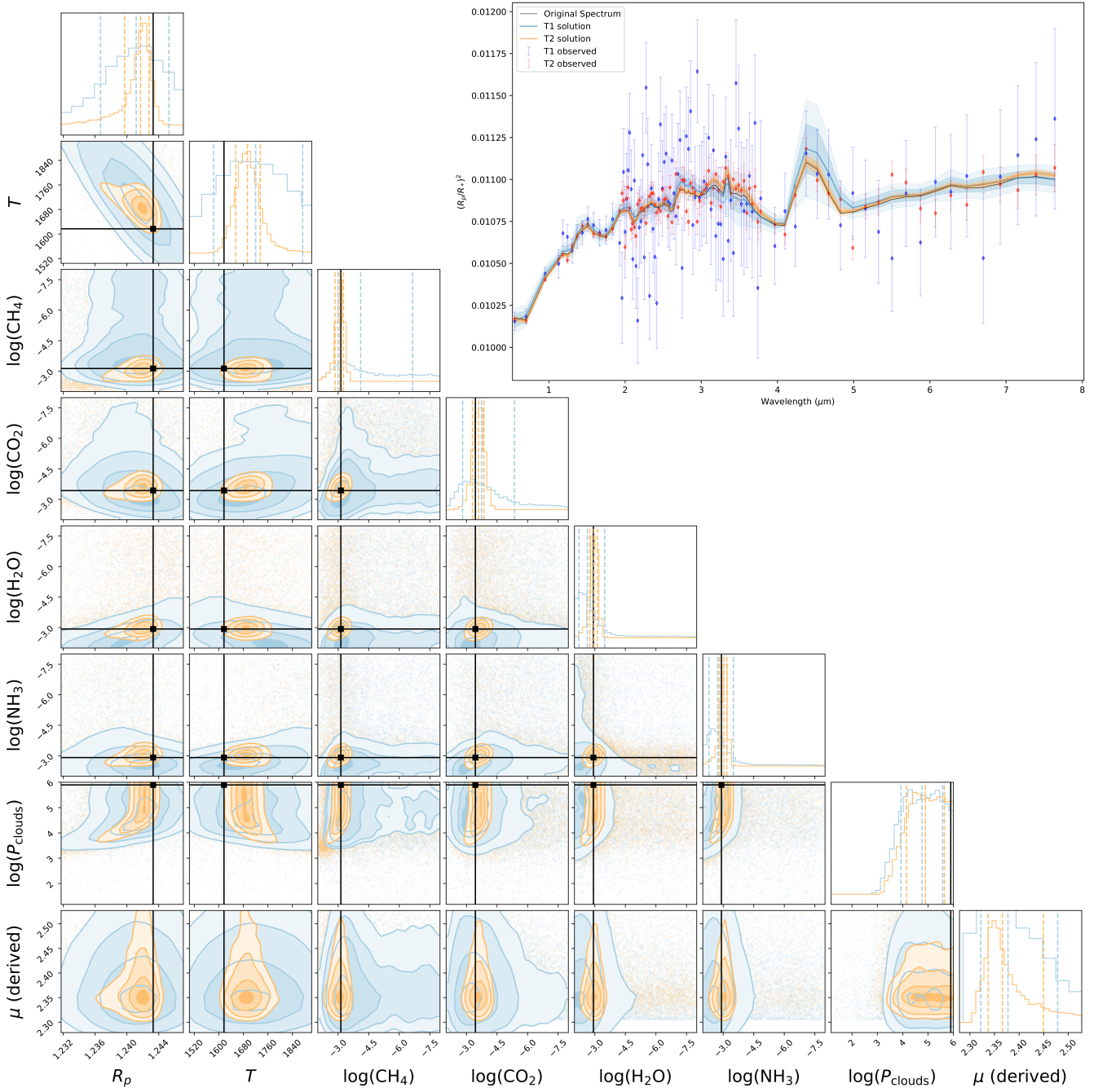}
	\caption{Retrieved spectra and posteriors. The corner plot shows the posteriors for each retrieved parameter using  Tier 1 (blue) and Tier 2 (orange) observed data. Input values are shown by the black lines. The panel in the top right corner shows the retrieved spectra from Tier 1 (blue) and Tier 2 data using coloured shaded bands for 1 and $2-\sigma$ uncertainties, and the input (black solid line). The notation $\log(X)$, where $X$ is one of CH$_4$, CO$_2$, H$_2$O or NH$_3$, represents the logarithm of the molecular abundance of the given species and should be compared to $\log(Ab_X)$.}
	\label{fig:retrieval}
\end{figure}

To perform the retrieval, we use TauREx 3 \citep{Al-Refaie2020}. The parameters fitted with fit boundaries, true and retrieved values are listed in Tab. \ref{tab:retrieval}, while the retrieved solutions and posteriors are shown in Fig. \ref{fig:retrieval}. 

For the selected planet, we notice that in Tier 2 the abundances of the molecules considered are well constrained, and, as expected, low level (high pressure) clouds are undetected in both cases. 


\begin{table}[]
	\centering
	\caption{Retrieval parameter table showing fit boundaries, true inputs, and retrieved parameters with uncertainties for Tier 1 and Tier 2 observations. As in Fig. \ref{fig:retrieval}, the notation $\log(X)$, where $X$ is one of CH$_4$, CO$_2$, H$_2$O or NH$_3$, represents the retrieved logarithm of the molecular abundance of the given species and should be compared to the input $\log(Ab_X)$.} 

	\label{tab:retrieval}
	\begin{tabular}{ccccc}
		\hline\noalign{\smallskip}
		{\bf \centering Name} & {\bf \centering Boundaries} & {\bf \centering True value}& {\bf \centering Tier 1 retrieved} & {\bf \centering Tier 2 retrieved} \\
		\noalign{\smallskip}\hline\noalign{\smallskip}   
		$R_p \; [R_{Jup}]$ & $[0.5 \to 2] ]$ & 1.24 & $1.241^{+0.004}_{-0.004}$ & $1.2412^{+0.0020}_{-0.0011}$ \\
		$T_p \; [K]$ & $[800 \to 2400]$ & 1617 & $1720^{+153}_{-137}$ & $1693^{+42}_{-38}$ \\
		$\log(CH_4)$ & $[-8 \to -2 ]$ & -3.13 & $-4.11^{+1.11}_{-2.55}$ & $-3.08^{+0.23}_{-0.18}$ \\
		$\log(CO_2)$ & $[-8 \to -2 ]$ & -3.44 & $-3.74^{+0.93}_{-1.61}$ & $-3.59^{+0.30}_{-0.26}$\\
		$\log(H_2O)$ & $[-8 \to -2 ]$ & -2.93 & $-2.63^{+0.42}_{-0.85}$ & $-2.96^{+0.20}_{-0.17}$ \\
		$\log(NH_3)$ & $[-8 \to -2 ]$ & -2.91 & $-2.73^{+0.43}_{-0.77}$ & $-3.03^{+0.22}_{-0.15}$ \\
		$\log(P_{clouds})$ & $[-3 \to 6]$ & 5.90 & $4.76^{+0.81}_{-0.83}$ & $4.89^{+0.75}_{-0.74}$ \\
		
		\noalign{\smallskip}\hline
	\end{tabular}
\end{table}

The Tier 1 results can be linked to our previous analysis on molecular input abundance detection (Sec. \ref{sec:relation_abundance}). We compute the probability to have molecular abundances greater than $10^{-4}$ from the retrieval posteriors and compare these with the probability obtained with our metric (Fig. \ref{fig:prob}). In this case, the measured $M_{mol}$ are: $M_{CH_4}=-0.47$, $M_{CO_2}=0.54$ and $M_{H_2O}=0.29$. The results are listed in Tab. \ref{tab:retrieval_prob}.  
Tier 2 observations provide a confident detection of methane, carbon dioxide and water, while Tier 1 retrievals are broadly comparable to our metric approach in detecting the presence of these molecules. 

These results appear to confirm that spectral retrievals may not be best suited or at the very least necessary to analyse Tier 1 data. Retrievals are model-dependent, and one needs to define planet parameters, as well as cross-sections, pressure-temperature profiles, etc. Priors might need to be imposed to ensure convergence. Retrievals are also computationally expensive, making it not trivial to conduct the analysis on hundreds of targets. A photometric metric instead, is model-independent, which may be an advantage when assessing a planet observation for the first time. The full analysis takes only minutes on a desktop computer to reduce 1000 observations. 



\begin{table}[]
	\centering
	\caption{Probability to have $Ab_{mol}> 10^{-4}$ for each molecule computed from $M_{mol}$ and from Tier 1 and Tier 2 retrieval posteriors. The numbers refer to the planet case discussed in Sec. \ref{sec:retrieval}.}
	\label{tab:retrieval_prob}
	\begin{tabular}{cccc}
		\hline\noalign{\smallskip}
		{\bf \centering Molecule} & {\bf \centering $M_{mol}$ $[\%]$}& {\bf \centering Tier 1 $[\%]$} & {\bf \centering Tier 2 $[\%]$} \\
		\noalign{\smallskip}\hline\noalign{\smallskip}   
		CH$_4$ & $49$ & $ 48 $ & $ 100 $ \\
		CO$_2$ & $78$ & $ 58 $ & $ 94 $\\
		H$_2$O & $56$ & $ 89 $ & $ 100 $ \\
		\noalign{\smallskip}\hline
	\end{tabular}
\end{table}

\subsection{Comparison with Deep and Machine Learning} \label{sec:deep_discussion}

ML techniques are difficult to interpret, and so a comparison between their performance and that of our metric can help us in gaining confidence in the outcomes from ML classifiers. For this purpose, we consider a planet as bearing a molecule if $Ab_{mol}>10^{-4}$. Then with our metric we select all planets that have $M_{CH_4}\geq 0.22$ that according to Fig. \ref{fig:prob} corresponds to a probability of $\sim 68.3\%$ to have a $Ab_{mol}>10^{-4}$ for CH$_4$.We repeat the same procedure, letting $M_{CO_2}\geq 0.26$ for CO$_2$ and $M_{H_2O}\geq 0.80$ for H$_2$O. In each sample, we check how many of the selected planets have molecular abundances in excess of $10^{-4}$,  obtaining a percentage of success for our metric (or metric precision). 
In the same way, we check how many of the planets flagged by each of the Deep and Machine Learning algorithms in the full sample actually bear the molecules, such that we can compare their precision performance in Tab. \ref{tab:Mmol-Knn_perf}.

We notice a marginally better success rate for Deep and Machine Learning algorithms in the cases of KNN and MLP, while RFC and SVC algorithms suggest a better performance when compared to that of the metric.  
Better performances are expected because, while our metric considers only specific bins in the spectrum, the classification algorithms gather information from all the spectral data points. The comparable performance of the metric with the KNN and MLP suggests that the molecular bands chosen for the metric are not far from ideal, but the comparatively better performances of RFC and SVC provide an indication that margins for improvement may exist.

While more work is required along this path, which is beyond the scope of this work, Deep and ML appear to be very promising for this classification problem, and we shall leave to dedicated works, as the one presented in \citet{yip2020}, a more exhaustive investigation of these techniques, their comparison with more physically motivated strategies similar to the metric, and a thorough investigation of biases that may affect all these techniques. 

\begin{table}[]
	\centering
	\caption{Percentages of positive detection for our metric, compared to Deep Learning algorithms precision. To assess the presence of a molecule we flag a planet if $Ab_{mol}>10^{-4}$. We investigate CH$_4$ in the first row, CO$_2$ in the second and H$_2$O in the third, selecting the planets with $M_{CH_4}\geq 0.22$ (first row), $M_{CO_2}\geq 0.26$ (second row) and $M_{H_2O}\geq 0.80$ (third row).}
	\label{tab:Mmol-Knn_perf}
	
	\begin{tabular}{cccccc}

		{\bf \centering Molecule} & {\bf \centering $M_{mol}$ [$\%$]} & {\bf \centering KNN [$\%$]}& {\bf \centering MLP[$\%$]} & {\bf \centering RFC[$\%$]} & {\bf \centering SVC[$\%$]} \\
		\noalign{\smallskip}\hline\noalign{\smallskip} 
		CH$_4$ & 69\footnote{These percentages arise from a discrete distribution of data and therefore we cannot exactly identify the $68.3\%$ quantity. In this case $69\%$ is the closest possible value.} & 75.4 & 84.2 & 92.5 & 90.1\\
		CO$_2$ & 68.3 & 71.4 & 75.8 & 83.1 & 83.5\\
		H$_2$O & 68.3 & 74.5 & 79.0 & 96.7 & 99.4 \\
		\noalign{\smallskip}\hline		
	\end{tabular}
\end{table}

\section{Conclusion}
\label{sec:conclusion}

This work presents data analysis methods to extract atmospheric information from \ARIEL\ Tier 1 observations of a large and diverse sample of exoplanets. 
\ARIEL's Tier 1 has been optimised as a reconnaissance survey of exoplanets, with SNR larger than 7 after averaging the observed spectra in about 7 photometric data points over the 0.5 -- 7.8 $\mu$m wavelength range. Therefore, having only 7 effective data points per spectrum, Tier 1 data may not be ideally suited for detailed spectral retrieval and to constrain chemical abundances, for which Tier 2 or 3 observations are needed.  However, Tier 1 data contain a wealth of information such as the spectral signatures of important molecules, whose presence can in principle be detected, therefore enabling targets to be classified, and can be used to assess planets with featureless spectra.  

In this work we simulate the entire population of exoplanets using Alfnoor, assigning a randomised atmosphere to each planet in the Ariel Mission Reference Sample that comprises a diverse population of 1000 exoplanetary targets. We consider primary atmospheres with contributions from clouds, methane, water, carbon dioxide and ammonia. This simulated data set is expected to be representative of the \ARIEL\ Tier 1 reconnaissance survey. 

The aim of this paper is threefold: (1) to show the capability of Tier 1 to detect featureless spectra, (2) to define a metric to classify and select planets to be re-observed in higher resolution Tiers and (3) to introduce other strategies that can be used to maximise the science exploitation of \ARIEL's Tier 1 data, for consideration in future studies.

(1) We presented a reliable method to identify flat spectra. By dividing the \ARIEL\ wavelength range into 4  bands, we classify as flat those planets where the 4 spectral bands response is compatible with a flat line, following a $\chi^2$ test.


(2) We developed a model-independent metric that bins the observed spectra over selected bands bearing the signatures of the molecules under investigation. From the observed spectrum alone, this method proves capable to indicate the presence of an atmosphere and its possible composition, independently of the planet parameters such as mass, size and temperature. 
Applying the metric to a Tier 1 observed spectrum, we find a $1-\sigma$ confidence level in identifying CH$_4$, CO$_2$ or H$_2$O when their abundance in the atmosphere is in  excess of  $10^{-4}$ in mixing ratio, and their estimates $M_{CH_4}\geq 0.22$, $M_{CO_2}\geq 0.26$ or $M_{H_2O}\geq 0.80$, respectively, demonstrating how the metric may be used in a statistically quantitative way.
However,  we find that the metric is  biased, and the bias depends on the magnitude of the instrumental noise. De-biasing the metric is required for its predictions to be quantitative. De-biasing is expected to be possible, following a detailed characterisation of the instrumental uncertainties,  and we reserve to investigate these aspects in a future study. 

The metric struggles to separate H$_2$O and NH$_3$. This may be partially due to the effect of a bias, or, more likely, because of the two molecules partially overlapping features. However, the metric is successful in classifying these targets as having an atmosphere. Should these targets be selected for Tier 2 observations, a spectra retrieval analysis can constrain all abundances to high significance. 


(3) We have performed a preliminary comparison of four different Deep and Machine Learning algorithms for the chemical classification of Tier 1 atmospheres. We find that their performance in identifying the presence of a certain molecule in the spectrum is marginally better than that of the metric in the case of KNN and MLP, but RFC and SVC outperform the metric, justifying a detailed follow-up study in future work. 


\appendix
\section{Analytical derivation of the metric} \label{sec:analytic_metric}

As mentioned in Sec. \ref{sec:metric}, the metric here presented is (i) sensitive to the molecules, (ii) independent from the planet size, and (iii) independent from the scale height. To show it, we start by using the following notation:  in transmission spectroscopy we are measuring
\begin{equation}
    \frac{\Delta f}{f}(\lambda) = \frac{ R_{pl}^2 + 2 R_{pl} \cdot z(\lambda)}{ R_{\star}^2} 
\end{equation}
where $f$ is the measured flux from the star, $\Delta f$ is the difference between the flux measured during the transit and the one measured out of transit, $R_\rho$ and $R_\star$ are the planet and the star radii respectively; $z(\lambda)$ is the measured wavelength dependent transit depth. 
Now, applying the definition of $z(\lambda)$ from \citet{Lecavelier2008A&A}, 
\begin{equation}
    z(\lambda) = H \ln \left( \frac{\epsilon_{abs}\sigma_{abs}(\lambda)P_0}{\tau_{eq}} \sqrt{\frac{2 \pi R_p H }{k_B^2 T_p^2}}\right)
\end{equation}
where $\epsilon_{abs}$ and $\sigma_{abs}$ are the abundance and cross section of the main absorbent $abs$ at the $\lambda$ wavelength. $H$ is the scale height to which correspond the $P_0$ pressure and the $\tau_{eq}$ is the equivalent optical depth. Therefore, we have
\begin{equation}
    \frac{\Delta f}{f}(\lambda) = \frac{ R_{pl}^2 + 2 R_{pl} H \ln \left( \frac{\epsilon_{abs}\sigma_{abs}(\lambda)P_0}{\tau_{eq}} \sqrt{\frac{2 \pi R_p H }{k_B^2 T_p^2}}\right)}{ R_{\star}^2} = \frac{ R_{pl}^2 + 2 R_{pl} H \cdot Z(\lambda)}{ R_{\star}^2} 
\end{equation}
where, for simplicity, we called 
\begin{equation}\label{eq:Z}
Z(\lambda) =\ln \left( \frac{\epsilon_{abs}\sigma_{abs}(\lambda)P_0}{\tau_{eq}} \sqrt{\frac{2 \pi R_p H }{k_B^2 T_p^2}}\right) 
\end{equation}

Therefore, to measure $S_{band_i}$ in eq. \ref{eq:sband} corresponds to computing the mean in the band:

\begin{equation}
    S_{band_i} = \left( \frac{\Delta f}{f} \right)_{band_i} = \frac{R_{pl}^2}{R_{\star}^2} + \frac{2 R_{pl}H}{ R_{\star}^2} \cdot \frac{\sum_j^M Z_j}{M} = \frac{R_{pl}^2}{R_{\star}^2} + \frac{2 R_{pl}H}{ R_{\star}^2}\cdot Z_{band_i}
\end{equation}
where $Z_j$ is the equivalent of eq. \ref{eq:Z} in the $j^\mathrm{th}$ spectral bin and $Z_{band_i} = \frac{\sum_j^M Z_j}{M}$. Therefore, the dispersion of eq. \ref{eq:sigma_sband} is computed as
\begin{equation}
    \sigma_{band_i} = \frac{2 R_{pl}H}{ R_{\star}^2} \sqrt{\frac{\sum_j^M (Z_j-Z_{band_i})^2}{M}} = \frac{2 R_{pl}H}{ R_{\star}^2}  \cdot \sigma_{Z_{band_i}}
\end{equation}
where $\sigma_{Z_{band_i}} = \sqrt{\frac{\sum_j^M (Z_j-Z_{band_i})^2}{M}}$.

By combining the previous equations as done in eq. \ref{eq:mmol}, we finally obtain
\begin{equation}
    M_{mol} = \frac{1}{N} \sum_i^N \frac{\frac{2 R_{pl}H}{ R_{\star}^2} \left( Z_{band_i}-Z_{norm}\right)}{\frac{2 R_{pl}H}{ R_{\star}^2} \sqrt{\sigma_{Z_{band_i}}^2 + \sigma_{Z_{norm}}^2}} = \frac{1}{N} \sum_i^N \frac{ Z_{band_i}-Z_{norm}}{ \sqrt{\sigma_{Z_{band_i}}^2 + \sigma_{Z_{norm}}^2}}
\end{equation}
Therefore, we remove the planet and star radii dependence in the measurement. 
Similarly to what has been done in \citet{desert2009ApJ}, the subtraction between $Z_{band_i}$ and $Z_{norm}$ finally removes the scale height dependency as 
\begin{equation}
    Z_{band_i} - Z_{norm} = \ln \left( \frac{\epsilon_{abs, \, band_i}\sigma_{abs, \,  band_i}}{\epsilon_{abs, norm}\sigma_{abs, norm}}\right)
\end{equation}
where $\epsilon_{abs, \, band_i}\sigma_{abs, \,  band_i}$ is the equivalent of $\epsilon_{abs} \sigma_{abs}(\lambda)$ in the band. This factor identifies the contribution of the main absorber in the band. Therefore, if we compare a band where a certain molecule has a strong feature, with one where is not supposed to give contributions to the spectrum, we can identify the molecular presence, compared to what is present in the second band. 

So, finally $M_{mol}$ becomes
\begin{equation}
    M_{mol} = \frac{1}{N} \sum_i^N \frac{ \ln \left( \frac{\epsilon_{abs, \, band_i}\sigma_{abs, \,  band_i}}{\epsilon_{abs, norm}\sigma_{abs, norm}}\right)}{ \sqrt{\sigma_{Z_{band_i}}^2 + \sigma_{Z_{norm}}^2}}
\end{equation}
So, as promised, the metric is also sensitive to the molecular content. 

To summarise, we removed the star, planet and atmosphere size dependencies by subtracting the interesting feature bands for a normalisation band and dividing the results by the combined dispersion. This results in a metric that is sensitive to the molecules contained in the atmosphere, but introduces a bias. 
In fact, the spectral dispersion $\sigma_{Z_{band_i}}$ depends on both the atmospheric feature dispersion and on the observational noise.

\acknowledgments
This work has been supported by ASI grant N. 2018.22.HH.O. and UCL Cities partnerships Programme. The project also received founding from the European Research Council (ERC) under the European Union's Horizon 2020 research and innovation programme (grant agreement No 758892, ExoAI) and under the European Union's Seventh Framework Programme (FP7/2007-2013)/ERC grant agreement No 617119 (ExoLights). Furthermore, we acknowledge funding by the Science and Technology Funding Council (STFC) grants ST/K502406/1, ST/P000282/1, ST/P002153/1 and ST/S002634/1. The authors acknowledge the contribution of the anonymous referees that greatly improved this work.

\clearpage

\bibliography{main} 
\bibliographystyle{aasjournal}

\end{document}